\documentclass[a4paper,11pt]{article}
\pdfoutput=1 

\usepackage{jheppub} 

\usepackage[T1]{fontenc} 
\usepackage{subcaption}

\makeatletter
\DeclareRobustCommand*{\bfseries}{%
   \not@math@alphabet\bfseries\mathbf
   \fontseries\bfdefault\selectfont
   \boldmath
}
\makeatother

\newcommand{\tr}{\text{tr}}

\usepackage{tikz}
\usepackage{tikz-3dplot}
\usepackage{verbatim}
\usetikzlibrary{decorations.pathreplacing}
\usepackage[framemethod=TikZ]{mdframed}
\usepackage{mathrsfs}

\usetikzlibrary{decorations.pathreplacing,decorations.markings}

\tikzset{
    >=stealth',
    punkt/.style={
           rectangle,
           rounded corners,
           draw=black, very thick,
           text width=6.5em,
           minimum height=2em,
           text centered},
    pil/.style={
           ->,
           thick,
           shorten <=2pt,
           shorten >=2pt,},
  on each segment/.style={
    decorate,
    decoration={
      show path construction,
      moveto code={},
      lineto code={
        \path [#1]
        (\tikzinputsegmentfirst) -- (\tikzinputsegmentlast);
      },
      curveto code={
        \path [#1] (\tikzinputsegmentfirst)
        .. controls
        (\tikzinputsegmentsupporta) and (\tikzinputsegmentsupportb)
        ..
        (\tikzinputsegmentlast);
      },
      closepath code={
        \path [#1]
        (\tikzinputsegmentfirst) -- (\tikzinputsegmentlast);
      },
    },
  },
  mid arrow/.style={postaction={decorate,decoration={
        markings,
        mark=at position .5 with {\arrow[#1]{stealth'}}
      }}}
}

 \newcommand{\ket}[1]{|#1\rangle}
\newcommand{\ketbra}[2]{|#1\rangle\!\langle#2|}

\newtheorem{theorem}{Theorem}

\newtheorem{conjecture}[theorem]{Conjecture}

\newtheorem{definition}[theorem]{Definition}

\newtheorem{principle}[theorem]{Principle}
\newenvironment{proof}[1][Proof]{\noindent\textbf{#1.} }{\ \rule{0.5em}{0.5em}}

\usetikzlibrary{decorations.pathmorphing}
\tikzset{snake it/.style={decorate, decoration=snake}}

\title{Quantum tasks in holography}

\author[a]{Alex May}

\affiliation[a]{The University of British Columbia}

\emailAdd{may@phas.ubc.ca}

\abstract{We consider an operational restatement of the holographic principle, which we call the principle of asymptotic quantum tasks. Asymptotic quantum tasks are quantum information processing tasks with inputs given and outputs required on points at the boundary of a spacetime. The principle of asymptotic quantum tasks states that tasks which are possible using the bulk dynamics should coincide with tasks that are possible using the boundary. We extract consequences of this principle for holography in the context of asymptotically AdS spacetimes. We argue for a novel connection between bulk causal structure and the phase transition in the boundary mutual information. Further, we note a connection between holography and quantum cryptography, where the problem of completing asymptotic quantum tasks has been studied earlier. We study the cryptographic and AdS/CFT approaches to completing asymptotic quantum tasks and consider the efficiency with which they replace bulk classical geometry with boundary entanglement.}

\begin{document} 
\maketitle
\flushbottom

\section{Introduction}

The holographic principle \cite{hooft1993dimensional,susskind1995world} asserts the dynamical equivalence of two theories, one defined on a $d+1$ dimensional spacetime $\mathcal{M}$ and the other on a $d$ dimensional spacetime, usually taken to be the boundary of $\mathcal{M}$. That this can ever occur is surprising, but the AdS/CFT correspondence \cite{maldacena1999large,witten1998anti} gives one concrete realization. In the context of AdS/CFT this dynamical equivalence is succinctly stated in terms of an equality of partition functions.

Another perspective one can take on this dynamical equivalence is an operational one, as an equivalence of what it is possible to accomplish in the bulk and in the boundary. To make this precise we need to make sense of the notion of doing the same task in the bulk and boundary, even though the bulk and boundary degrees of freedom may look very different. Our answer to this will be the notion of an \emph{asymptotic quantum task}, which in the bulk is stated in terms of inputs that come in from and outputs that go out to the spacetime boundary. Our rephrasing of the holographic principle is that asymptotic quantum tasks are possible in the boundary if and only if they are possible in the bulk. Applied to asymptotically AdS spacetimes, we find that this principle can be used to arrive at precise consequences. 

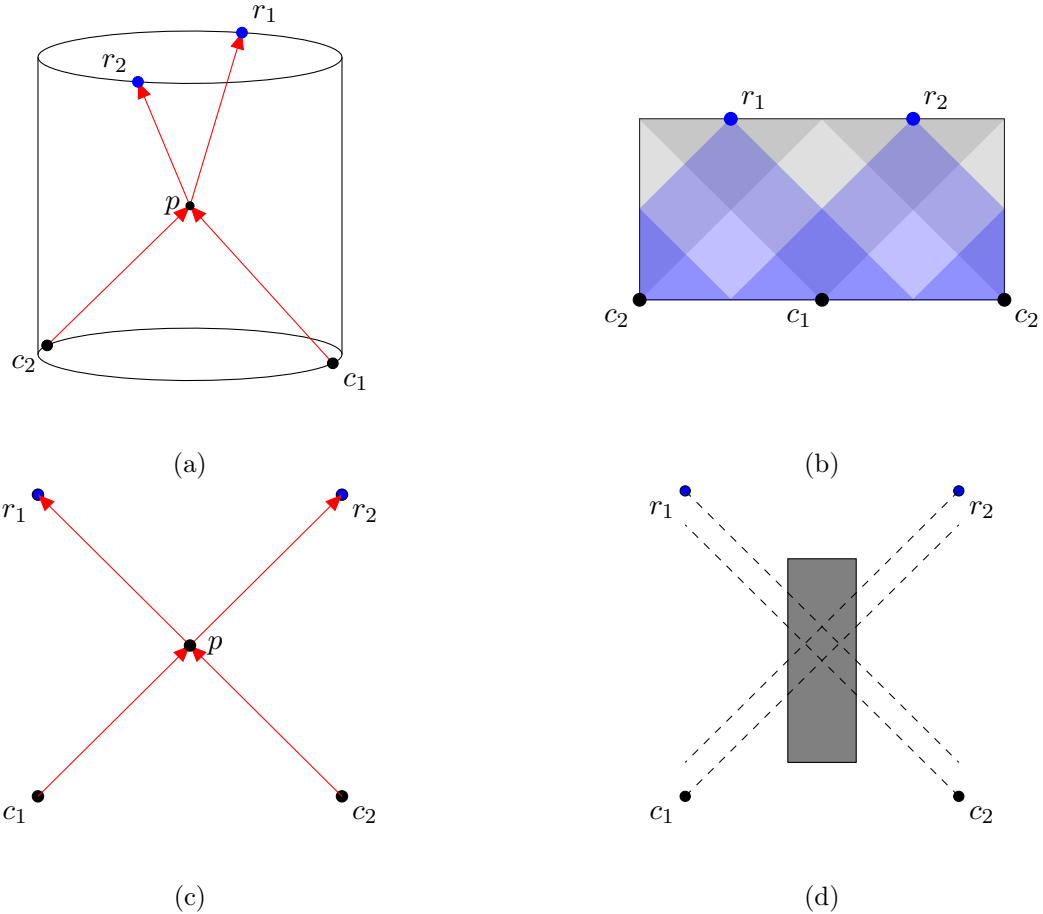
\begin{figure}
\begin{center}
\begin{subfigure}[b]{0.45\textwidth}
\begin{center}
\tdplotsetmaincoords{10}{0}
    \begin{tikzpicture}[scale=1,tdplot_main_coords]
    \tdplotsetrotatedcoords{0}{20}{0}
    \draw (-2,0,0) -- (-2,4,0);
    \draw (2,0,0) -- (2,4,0);
    
    \begin{scope}[tdplot_rotated_coords]
    \begin{scope}[canvas is xz plane at y=0]
    \draw (0,0) circle [radius=2];
    \end{scope}
    
    \begin{scope}[canvas is xz plane at y=4]
    \draw (0,0) circle [radius=2];
    \end{scope}
    
    \draw[red,-triangle 45] (2,0,0) -- (0,2,0);
    \draw[red,-triangle 45] (-2,0,0) -- (0,2,0);
    \draw[red,-triangle 45] (0,2,0) -- (0,4,2);
    \draw[red,-triangle 45] (0,2,0) -- (0,4,-2);
    
    \draw plot [mark=*, mark size=2] coordinates{(2,0,0)};
    \node[below right] at (2,0,0) {$c_1$};
    
    \draw plot [mark=*, mark size=2] coordinates{(-2,0,0)};
    \node[below left] at (-2,0,0) {$c_2$};
    
    \draw plot [mark=*, mark size=1.5] coordinates{(0,2,0)};
    \node[left] at (0,2,0) {$p$};
    
    \draw[blue] plot [mark=*, mark size=2] coordinates{(0,4,-2)};
    \node[above left] at (0,4,-2) {$r_2$};
    
    \draw[blue] plot [mark=*, mark size=2] coordinates{(0,4,2)};
    \node[above right] at (0,4,2) {$r_1$};
    
    \end{scope}

\end{tikzpicture}
\end{center}
\caption{}
\label{subfig:cylinder}
\end{subfigure}
\hfill
\begin{subfigure}[b]{.45\textwidth}
\begin{center}
\begin{tikzpicture}[scale=1.2]

 \draw (-2,0) -- (2,0) -- (2,2) -- (-2,2) -- (-2,0);
    
    \draw[gray,fill=gray,opacity=0.25] (-2,2) -- (0,0) -- (2,2);
    
    \draw[gray,fill=gray,opacity=0.25] (-2,2) -- (-2,0) -- (0,2) -- (-2,2);
    
    \draw[gray,fill=gray,opacity=0.25] (2,2) -- (2,0) -- (0,2) -- (2,2);
    
    \draw[blue,fill=blue,opacity=0.25] (-1,2) -- (-2,1) -- (-2,0) -- (1,0) -- (-1,2);
    
    \draw[blue,fill=blue,opacity=0.25] (1,2) -- (2,1) -- (2,0) -- (-1,0) -- (1,2);
    
    \draw[blue,fill=blue,opacity=0.25] (2,1) -- (2,0) -- (1,0) -- (2,1);
    \draw[blue,fill=blue,opacity=0.25] (-2,1) -- (-2,0) -- (-1,0) -- (-2,1);
    
    \draw[black] plot [mark=*, mark size=2] coordinates{(-2,0)};
    \node[below left] at (-2,0) {$c_2$};
    
    \draw[black] plot [mark=*, mark size=2] coordinates{(2,0)};
    \node[below right] at (2,0) {$c_2$};
    
    \draw[black] plot [mark=*, mark size=2] coordinates{(0,0)};
    \node[below left] at (0,0) {$c_1$};
    
    \draw[blue] plot [mark=*, mark size=2] coordinates{(-1,2)};
    \node[above right] at (-1,2) {$r_1$};
    
    \draw[blue] plot [mark=*, mark size=2] coordinates{(1,2)};
    \node[above right] at (1,2) {$r_2$};
    
    \node at (0,-1) {$ $};

\end{tikzpicture}

\end{center}
\caption{}
\label{subfig:boundary}
\end{subfigure}
\begin{subfigure}[b]{.45\textwidth}
\begin{center}
\begin{tikzpicture}[scale=0.5]
    
    \node[below left] at (-4,0) {$c_1$};
    \draw[fill=black] (-4,0) circle (0.15);

    \node[below right] at (4,0) {$c_2$};
    \draw[fill=black] (4,0) circle (0.15);

    \node[below right] at (4,8) {$r_2$};
    \draw[fill=blue] (4,8) circle (0.15);

    \node[below left] at (-4,8) {$r_1$};
    \draw[fill=blue] (-4,8) circle (0.15);
    
    \draw[red,-triangle 45] (-4,0) -- (0,4);
    \draw[red,-triangle 45] (4,0) -- (0,4);
    \draw[red,-triangle 45] (0,4) -- (4,8);
    \draw[red,-triangle 45] (0,4) -- (-4,8);
    
    \node[right] at (0.2,4) {$p$};
    \draw[fill=black] (0,4) circle (0.15);

\end{tikzpicture}
\end{center}
\caption{}
\end{subfigure}
\hfill
\begin{subfigure}[b]{.45\textwidth}
\begin{center}
\begin{tikzpicture}[scale=0.45]
    
    \node[below left] at (-4,0) {$c_1$};
    \draw[fill=black] (-4,0) circle (0.15);

    \node[below right] at (4,0) {$c_2$};
    \draw[fill=black] (4,0) circle (0.15);

    \node[below right] at (4,9) {$r_2$};
    \draw[fill=blue] (4,9) circle (0.15);

    \node[below left] at (-4,9) {$r_1$};
    \draw[fill=blue] (-4,9) circle (0.15);
    
    \draw[fill=gray] (-1,1) -- (1,1) -- (1,7) -- (-1,7) -- (-1,1);

    \draw[dashed] (-4,0) -- (4,8);
    \draw[dashed] (4,0) -- (-4,8);
    \draw[dashed] (4,9) -- (-4,1);
    \draw[dashed] (-4,9) -- (4,1);

\end{tikzpicture}
\end{center}
\caption{}
\label{subfig:tagging}
\end{subfigure}
\caption{a) An asymptotic quantum task in AdS space, shown from the bulk perspective. Alice receives quantum or classical systems $A_1$ at $c_1$ and $A_2$ at $c_2$. She must apply some quantum channel $\mathcal{N}:A_1A_2\rightarrow B_1B_2$ before returning the systems $B_1$ at $r_1$ and $B_2$ at $r_2$. An Alice living in the bulk may complete the task by bringing the inputs to the central point $p\in P \equiv J^+(c_1) \cap J^+(c_2) \cap J^-(r_1) \cap J^-(r_2)$, performing the needed operation, and sending the outputs to the appropriate $r_i$. b) According to the principle of asymptotic quantum tasks, Alice in the boundary must be able to complete the same task. However, the central region $P$ is empty when considered in the boundary geometry, as indicated by the shaded light cones. Nonetheless, the boundary theory must be able to accomplish the same task. c) A quantum task considered in the context of cryptography. The central region $P$ is available, analogous to the bulk setting shown in a). d) A quantum task with an excluded region, shown in grey. Alice will try to complete the quantum task without performing quantum operations within the grey region. The situation is analogous to the boundary perspective shown in b), since no central region is available. Such excluded regions were studied in the context of quantum tagging \cite{kent2006tagging}, where it was found that a central region could be replaced if and only if entanglement is distributed across it \cite{buhrman2014position}.}
\label{fig:cylinder1}
\end{center}
\end{figure}

As an initial example of an asymptotic quantum task consider the geometry of figure \ref{subfig:cylinder}. A quantum task with two input points $c_1,c_2$ and two output points $r_1,r_2$ has been arranged. Alice will receive quantum systems $A_1$ at $c_1$ and $A_2$ at $c_2$, and must apply a quantum channel $\mathcal{N}_{A_1A_2\rightarrow B_1B_2}$ before returning $B_1$ at $r_1$ and $B_2$ at $r_2$. In general the quantum channel will not be product, $\mathcal{N}_{A_1A_2\rightarrow B_1B_2}\neq \mathcal{N}_{A_1\rightarrow B_1}\otimes \mathcal{N}_{A_2\rightarrow B_2}$, and naively one expects that the inputs $A_1A_2$ must be brought together for the channel to be applied. More precisely, we expect that the channel must be applied somewhere in the region $J^+(c_1)\cap J^+(c_2)$\footnote{By $J^+(p)$ we mean all those points $q$ such that there is a causal curve connecting $p$ to $q$, and by $J^-(p)$ we mean all those points $q$ such that there is a causal curve from $q$ to $p$.}, which is the intersection of the future light cones of the input points. Further, we need to bring the outputs from applying this quantum channel to the points $r_1$ and $r_2$ so that we expect the channel must be applied in the region $P\equiv J^+(c_1) \cap J^+(c_2) \cap J^-(r_1) \cap J^-(r_2)$. Notice however that $c_1,c_2,r_1,r_2$ can be arranged such that in the bulk geometry $J^+(c_1) \cap J^+(c_2) \cap J^-(r_1) \cap J^-(r_2)$ is nonempty, while it is empty in the boundary geometry, as we show in figure \ref{subfig:boundary}\footnote{This geometric statement has appeared earlier, for example in \cite{maldacena2017looking}.}. The holographic principle then tells us that the boundary theory must be able to implement the channel  $\mathcal{N}_{A_1A_2\rightarrow B_1B_2}$, but somehow without making use of the central region. 

AdS/CFT provides one procedure by which this channel can be implemented in the boundary theory, despite the lack of central region. Intriguingly, this same problem of completing this task without access to a central region has arisen elsewhere, in the context of a quantum cryptographic problem called quantum tagging\footnote{This term should remind you of a graffiti artists signature, which proves they visited its location.} \cite{kent2011quantum}. In a quantum tagging scheme one party, call him Bob, tries to verify that another party, call her Alice, is performing quantum operations within a certain spacetime region $\mathcal{R}$. We illustrate such a quantum tagging scheme in figure \ref{subfig:tagging}. The scheme is a quantum task, consisting of input and output points and a certain quantum operation Alice is required to perform. General results \cite{buhrman2014position} show that Alice may always replace performing operations within $\mathcal{R}$ with entanglement distributed across $\mathcal{R}$. 

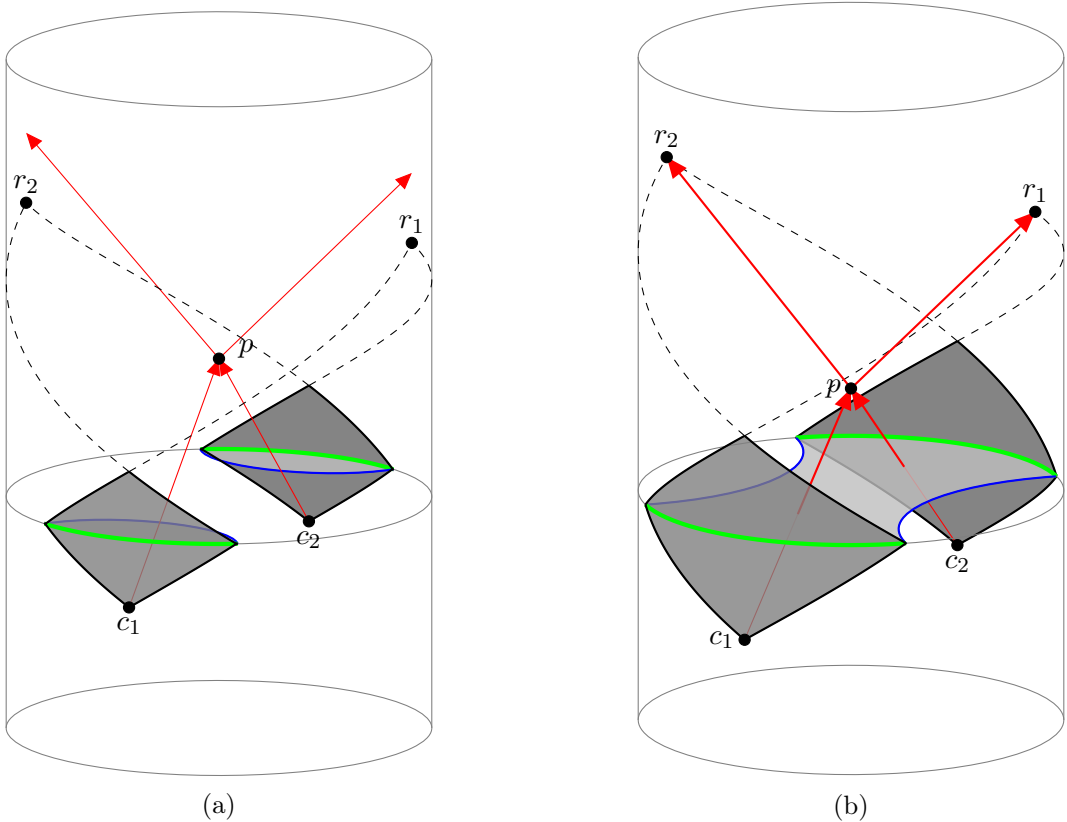
\begin{figure}
    \centering
    \begin{subfigure}{.45\textwidth}
    \centering
    
    \tdplotsetmaincoords{13}{0}
    \begin{tikzpicture}[scale=1.4,tdplot_main_coords]
    \tdplotsetrotatedcoords{0}{25}{0}
    \draw[gray] (-2,-0.25,0) -- (-2,6.25,0);
    \draw[gray] (2,-0.25,0) -- (2,6.25,0);
    \begin{scope}[tdplot_rotated_coords]
    
    \draw[domain=0:30,variable=\x,smooth,fill=black!60!,opacity=0.8] plot ({2*sin(\x)}, {1.34+\x/45}, {2*cos(\x)}) -- plot ({2*sin(30-\x)}, {2.66-(30-\x)/45}, {2*cos(30-\x)}) -- plot ({-2*sin(\x)}, {2.66-\x/45}, {2*cos(\x)}) -- plot ({-2*sin(30-\x)}, {1.34+(30-\x)/45}, {2*cos(30-\x)});
    
    \begin{scope}[canvas is xz plane at y=-0.25]
    \draw[gray] (0,0) circle[radius=2] ;
    \end{scope}
    
    \begin{scope}[canvas is xz plane at y=6.25]
    \draw[gray] (0,0) circle[radius=2] ;
    \end{scope}
    
    \begin{scope}[canvas is xz plane at y=2]
    
    \draw[gray] (0,0) circle (2);
    
    \draw [domain=60:120,fill=lightgray,opacity=0.8] plot ({2*cos(\x)}, {2*sin(\x)}) -- (-1, 1.73) to [out=-60,in=-120] (1, 1.73);
     \draw [domain=-60:-120,fill=lightgray,opacity=0.8] plot ({2*cos(\x)}, {2*sin(\x)}) -- (-1, -1.73) to [out=60,in=120] (1, -1.73);
    
    \draw [green,ultra thick,domain=60:120] plot ({2*cos(\x)}, {2*sin(\x)});
    
    \draw[blue, thick] (1, 1.73) to [out=-120,in=-60] (-1, 1.73);
    \draw[blue, thick] (1, -1.73) to [out=120,in=60] (-1, -1.73);
    
    \end{scope}
    
    \draw[domain=0:30,variable=\x,smooth,thick] plot ({2*sin(\x)}, {1.34+\x/45}, {2*cos(\x)});
    \draw[domain=0:30,variable=\x,smooth,thick] plot ({-2*sin(\x)}, {1.34+\x/45}, {2*cos(\x)});
    \draw[domain=0:30,variable=\x,smooth,thick] plot ({2*sin(\x)}, {2.66-\x/45}, {2*cos(\x)});
    \draw[domain=0:30,variable=\x,smooth,thick] plot ({-2*sin(\x)}, {2.66-\x/45}, {2*cos(\x)});
    
    \draw[red,-triangle 45] (0,1.34,2) -- (0,3.34,0);
    \draw[red,-triangle 45] (0,1.34,-2) -- (0,3.34,0);
    
    \draw[red,-triangle 45] (0,3.34,0) -- (2,5.34,0);
    \draw[red,-triangle 45] (0,3.34,0) -- (-2,5.34,0);
    
    \draw[domain=0:90,variable=\x,smooth,dashed] plot ({2*sin(\x)}, {2.66+\x/45}, {-2*cos(\x)});
    \draw[domain=0:90,variable=\x,smooth,dashed] plot ({2*sin(\x)}, {2.66+\x/45}, {2*cos(\x)});
    
    \draw[domain=0:90,variable=\x,smooth,dashed] plot ({-2*sin(\x)}, {2.66+\x/45}, {-2*cos(\x)});
    \draw[domain=0:90,variable=\x,smooth,dashed] plot ({-2*sin(\x)}, {2.66+\x/45}, {2*cos(\x)});
    
    \draw[domain=0:30,variable=\x,smooth,fill=black!50!,opacity=0.8] plot ({2*sin(\x+180)}, {1.34+\x/45}, {2*cos(\x+180)}) -- plot ({2*sin(180+30-\x)}, {2.66-(30-\x)/45}, {2*cos(180+30-\x)}) -- plot ({-2*sin(180+\x)}, {2.66-\x/45}, {2*cos(180+\x)}) -- plot ({-2*sin(180+30-\x)}, {1.34+(30-\x)/45}, {2*cos(180+30-\x)});
    
    \draw plot [mark=*, mark size=1.5] coordinates{(0,1.34,2)};
    \node[below] at (0,1.34,2) {$c_2$};
    \draw plot [mark=*, mark size=1.5] coordinates{(0,1.34,-2)};
    \node[below] at (0,1.34,-2) {$c_1$};
    \draw plot [mark=*, mark size=1.5] coordinates{(0,3.34,0)};
    \node[right] at (0.1,3.45,0) {$p$};
    
    \draw plot [mark=*, mark size=1.5] coordinates{(2,4.66,0)};
    \node[above] at (2,4.66,0) {$r_1$}; 
    \draw plot [mark=*, mark size=1.5] coordinates{(-2,4.66,0)};
    \node[above] at (-2,4.66,0) {$r_2$}; 
    
    \begin{scope}[canvas is xz plane at y=2]
    \draw [green,ultra thick,domain=-60:-120] plot ({2*cos(\x)}, {2*sin(\x)});
    \end{scope}
    
    \draw[domain=0:30,variable=\x,smooth,thick] plot ({2*sin(\x+180)}, {1.34+\x/45}, {2*cos(\x+180)});
    \draw[domain=0:30,variable=\x,smooth,thick] plot ({-2*sin(\x+180)}, {1.34+\x/45}, {2*cos(\x+180)});
    \draw[domain=0:30,variable=\x,smooth,thick] plot ({2*sin(\x+180)}, {2.66-\x/45}, {2*cos(\x+180)});
    \draw[domain=0:30,variable=\x,smooth,thick] plot ({-2*sin(\x+180)}, {2.66-\x/45}, {2*cos(\x+180)});
    
    \end{scope}
    \end{tikzpicture}
    \caption{}
    \end{subfigure}
    \hfill
    \begin{subfigure}{.45\textwidth}
    \centering
    \tdplotsetmaincoords{15}{0}
    \begin{tikzpicture}[scale=1.4,tdplot_main_coords]
    \tdplotsetrotatedcoords{0}{30}{0}
    \draw[gray] (-2,-0.25,0) -- (-2,6.25,0);
    \draw[gray] (2,-0.25,0) -- (2,6.25,0);
    
    \begin{scope}[tdplot_rotated_coords]
    
    \draw[domain=0:45,variable=\x,smooth, fill=black!60!,opacity=0.8] plot ({-2*sin(\x)}, {1+\x/45}, {2*cos(\x)}) -- plot ({-2*sin((45-\x))}, {3-(45-\x)/45}, {2*cos(45-\x)}) --  plot ({2*sin(\x)}, {3-\x/45}, {2*cos(\x)}) -- plot ({2*sin(45-\x)}, {1+(45-\x)/45}, {2*cos(45-\x)});
    
    \draw[domain=0:45,variable=\x,smooth,thick] plot ({-2*sin(\x)}, {1+\x/45}, {2*cos(\x)});
    \draw[domain=0:45,variable=\x,smooth,thick] plot ({2*sin(\x)}, {1+\x/45}, {2*cos(\x)});
    \draw[domain=0:45,variable=\x,smooth,thick] plot ({-2*sin(\x)}, {3-\x/45}, {2*cos(\x)});
    \draw[domain=0:45,variable=\x,smooth,thick] plot ({2*sin(\x)}, {3-\x/45}, {2*cos(\x)});
    
    \begin{scope}[canvas is xz plane at y=-0.25]
    \draw[gray] (0,0) circle[radius=2] ;
    \end{scope}
    
    \begin{scope}[canvas is xz plane at y=6.25]
    \draw[gray] (0,0) circle[radius=2] ;
    \end{scope}
    
    \draw[red] (0,1,-2) -- (0,2,-1);
    \draw[red] (0,1,2) -- (0,2,1);
    
    \begin{scope}[canvas is xz plane at y=2]
    
    \draw[gray] (0,0) circle (2);
    
    \draw [domain=-45:45,fill=lightgray,opacity=0.8] plot ({2*cos(\x+90)}, {2*sin(\x+90)}) -- (-1.41,1.41) to [out=-45,in=45] (-1.41,-1.41) -- plot ({2*cos(\x-90)}, {2*sin(\x-90)}) -- (1.41,-1.41) to [out=135,in=-135] (1.41,1.41);
    
    \draw [green,ultra thick,domain=-45:45] plot ({2*cos(\x+90)}, {2*sin(\x+90)});
    
    \draw[blue,thick] (1.41,1.41) to [out=-135,in=+135] (1.41,-1.41);
    \draw[blue,thick] (-1.41,1.41) to [out=-45,in=45] (-1.41,-1.41);
    
    \end{scope}
    
    \draw[domain=0:90,variable=\x,smooth,dashed] plot ({2*sin(\x+180)}, {3+\x/45}, {2*cos(\x+180)});
    \draw[domain=0:90,variable=\x,smooth,dashed] plot ({2*sin(\x+180)}, {3+\x/45}, {-2*cos(\x+180)});
    \draw[domain=0:90,variable=\x,smooth,dashed] plot ({-2*sin(\x+180)}, {3+\x/45}, {2*cos(\x+180)});
    \draw[domain=0:90,variable=\x,smooth,dashed] plot ({-2*sin(\x+180)}, {3+\x/45}, {-2*cos(\x+180)});
    
    \draw[thick, red,-triangle 45] (0,3,0) -- (2,5,0);
    \draw[thick, red,-triangle 45] (0,3,0) -- (-2,5,0);
    
    \draw[thick,red,-triangle 45] (0,2,-1) -- (0,3,0);
    \draw[thick,red,-triangle 45] (0,2,1) -- (0,3,0);
    
    \draw plot [mark=*, mark size=1.5] coordinates{(2,5,0)};
    \node[above] at (2,5,0) {$r_1$};
    \draw plot [mark=*, mark size=1.5] coordinates{(-2,5,0)};
    \node[above] at (-2,5,0) {$r_2$};
    
    \draw[domain=0:45,variable=\x,smooth, fill=black!50!,opacity=0.8] plot ({-2*sin(\x+180)}, {1+\x/45}, {2*cos(\x+180)}) -- plot ({-2*sin((45-\x)+180)}, {3-(45-\x)/45}, {2*cos(45-\x+180)}) --  plot ({2*sin(\x+180)}, {3-\x/45}, {2*cos(\x+180)}) -- plot ({2*sin(45-\x+180)}, {1+(45-\x)/45}, {2*cos(45-\x+180)});
    
    \begin{scope}[canvas is xz plane at y=2]
    \draw [green,ultra thick,domain=-45:45] plot ({2*cos(\x-90)}, {2*sin(\x-90)});
    \end{scope}
    
    \draw[domain=0:45,variable=\x,smooth,thick] plot ({-2*sin(\x+180)}, {1+\x/45}, {2*cos(\x+180)});
    \draw[domain=0:45,variable=\x,smooth,thick] plot ({2*sin(\x+180)}, {1+\x/45}, {2*cos(\x+180)});
    \draw[domain=0:45,variable=\x,smooth,thick] plot ({-2*sin(\x+180)}, {3-\x/45}, {2*cos(\x+180)});
    \draw[domain=0:45,variable=\x,smooth,thick] plot ({2*sin(\x+180)}, {3-\x/45}, {2*cos(\x+180)});
    
    \draw plot [mark=*, mark size=1.5] coordinates{(0,1,-2)};
    \node[left] at (0,1,-2) {$c_1$};
    \draw plot [mark=*, mark size=1.5] coordinates{(0,1,2)};
    \node[below] at (0,1,2) {$c_2$};
    \draw plot [mark=*, mark size=1.5] coordinates{(0,3,0)};
    \node[left] at (0,3,0) {$p$};
    
    \end{scope}
    \end{tikzpicture}
    \caption{}
    \end{subfigure}
    \caption{Combining our requirement on asymptotic quantum tasks, insights from quantum cryptography, and the Ryu-Takayangi formula we arrive at a precise result that applies to asymptotically AdS spacetimes with holographic descriptions: whenever the minimal surface enclosing $R_1R_2$ is given by the union of the minimal surfaces separately enclosing $R_1$ and $R_2$, the intersection of four lightcones becomes non-empty. The regions $R_1$ and $R_2$ are shown in green. Their domain of dependence defines the points $c_1$ and $c_2$. Shooting null rays (dashed lines) from $R_1$ and $R_2$ defines $r_1$ and $r_2$. Then $\mathcal{A}(R_1R_2) = \mathcal{A}(R_1)\cup \mathcal{A}(R_2)$ implies that $J^-(r_1)\cap J^-(r_2) \cap J^+(c_1)\cap J^+(c_2) = \emptyset$, as occurs at left. When $\mathcal{A}(R_1R_2) \neq \mathcal{A}(R_1)\cup \mathcal{A}(R_2)$ then we may find that the light cone intersection becomes non-empty. The implication is if and only if in the case of vacuum AdS. Details are given in section \ref{sec:entanglement-geometry}.}
    \label{fig:phasetransition3d}
\end{figure}

In both AdS/CFT and quantum tagging it is entanglement that replaces the use of the bulk central point. In fact, in the context of cryptography it has been shown that completing the task in figure \ref{subfig:tagging} without entering the grey region is impossible unless entanglement is available \cite{buhrman2014position}. In the context of holography we can leverage this proof to argue boundary regions must be entangled whenever a set of four spacetime points constructed from the regions have a central point. We give this as the following conjecture.\\

\noindent \textbf{Conjecture \ref{conj:entangledregions}}
\emph{Consider four spacetime points $c_1,c_2,r_1$ and $r_2$. Then if the central region $J^-(r_1)\cap J^-(r_2) \cap J^+(c_1)\cap J^+(c_2)$ considered in the bulk geometry is non-empty while the boundary central region is empty, then we have that $I(R_1:R_2)$ is $O(N^2)$, where the regions $R_i$ are defined according to $D(R_i) = \hat{J}^+(c_i)\cap \hat{J}^-(r_1)\cap \hat{J}^-(r_2)$.}\\

\noindent If we now assume the Ryu-Takayanagi formula, we can combine this result with the minimal surface prescription for calculating entanglement entropy \cite{ryu2006holographic,hubeny2007covariant} to arrive at a purely geometric statement.\\

\noindent \textbf{Conjecture \ref{conj:geometricversion}} 
Consider two disjoint boundary regions $R_1$ and $R_2$ defined by $D(R_i) = \hat{J}^+(c_i)\cap \hat{J}^-(r_1)\cap \hat{J}^-(r_2)$. Then if the minimal surface enclosing $R_1R_2$, denoted $\mathcal{A}(R_1R_2)$, is equal to $\mathcal{A}(R_1)\cup \mathcal{A}(R_2)$ then $J^-(r_1)\cap J^-(r_2) \cap J^+(c_1)\cap J^+(c_2) = \emptyset$.\\

\noindent This is illustrated in figure \ref{fig:phasetransition3d}. The argument is given in the main text, and supporting explicit checks in appendix \ref{app:minsurfaceandbulkpoint}.\footnote{Since the initial appearance of this article, a proof of this statement has appeared in \cite{may2019holographic}.} 

To argue for these statements we make use of the connection, highlighted in figure \ref{fig:cylinder1}, between quantum tagging and holography. We can also view this connection more broadly. Both the cryptographic protocols for spoofing quantum tagging schemes and AdS/CFT provide procedures for completing asymptotic quantum tasks from a boundary perspective. We refer to any such method as a \emph{holographic procedure}. Such procedures provide a method for replacing a bulk geometry with boundary entanglement. It is interesting to study how efficiently this can be done. We consider tasks that can be completed perfectly in the presence of a bulk classical geometry, and study how well they can be completed in the presence of a finite amount of entanglement. We can characterize the distance between a perfect completion of the task implemented by a channel $\mathcal{N}_{\text{ideal}}$ and an approximate completion of the task implemented by some channel $\mathcal{N}$ using the diamond-norm distance on channels \cite{kitaev2002classical,wilde2013quantum}. We find that this distance goes to zero with $1/I^{1/2}$ for the cryptographic procedure and $1/I^{1/4}$ for the AdS/CFT procedure, with $I$ a mutual information between two spatial regions relevant to the task.

This article is organized as follows. In section \ref{sec:QT} we set up the framework of quantum tasks and give some simple examples. Section \ref{sec:implications} uses the criterion that asymptotic quantum tasks be possible in the boundary when they are possible in the bulk to deduce some basic features of holographic theories dual to AdS spacetimes. Section \ref{sec:procedures} describes the AdS/CFT and cryptographic holographic procedures, and studies how efficiently they replace bulk geometry with boundary entanglement. We conclude with a discussion and comments on future directions in section \ref{sec:discussion}.

We summarize our notation here for reference. Upper case letters from the beginning of the alphabet $A_1, B_1,...$ will denote quantum systems, while lowercase letters $p, q, c_1,r_1,...$ will denote spacetime events. By $p\prec q$ we mean that there is a causal curve from event $p$ to event $q$. $J^+(p)\equiv\{q : p\prec q \}$ and $J^-(p)\equiv\{q: q\prec p \}$ denote the causal future and past of the event $p$. We will add hats to denote boundary regions, so that $J^+(p)$ is all those points in the bulk spacetime which are in the causal future of $p$, while $\hat{J}^+(p)$ considers only points in the boundary spacetime. Upper case letters from the middle of the alphabet $R_1,R_2,...$  will denote boundary spatial regions. The domain of dependence of a boundary region will be denoted $D(R)$, its causal wedge $\mathcal{C}(R)$, and its entanglement wedge $\mathcal{E}(R)$. Since it introduces no ambiguity, we will use the region itself or its domain of dependence to determine the causal and entanglement wedge, so that $\mathcal{C}(R) = \mathcal{C}(D(R))$ and $\mathcal{E}(R)=\mathcal{E}(D(R))$. 

\section{Asymptotic quantum tasks and holographic procedures}\label{sec:QT}

We begin by recalling \cite{kent2012quantum} what is meant by a relativistic quantum task. 
\begin{definition}
A \textbf{relativistic quantum task} is defined by a tuple \\ $\mathbf{T}=\{\mathcal{M},\ket{\Psi},\mathscr{A},\mathscr{B},\mathrm{c},\mathrm{r}, \mathcal{N}_{\mathscr{A}\rightarrow \mathscr{B}}\}$, where:
\begin{itemize}
    \item $\mathcal{M}$ is the spacetime in which the task occurs, it is described by a metric $g$ and ranges for the coordinates of that metric.
    \item $\ket{\Psi}$ is the state of any quantum fields present on some initial data time slice. This time slice should have all of the input and output points in its future domain of dependence.
    \item $\mathscr{A}=A_1A_2...A_n$ is the collection of all the input quantum systems and $\mathscr{B}=B_1...B_n$ is the collection of all the output quantum systems.
    \item $\mathrm{c}$ is the set of input points $\mathrm{c}=\{c_1,...,c_n\}$ and $\mathrm{r}$ is the set of output points $\mathrm{r}=\{r_1,...,r_n\}$.
    \item $\mathcal{N}_{\mathscr{A}\rightarrow \mathscr{B}}$ is a quantum channel that maps the input systems $\mathscr{A}$ to the output systems $\mathscr{B}$.
\end{itemize}
Alice receives the $A_i$ at the corresponding $c_i$ and must return the $B_i$ to the corresponding $r_i$, with the inputs and output states related by the quantum channel $\mathcal{N}_{\mathscr{A}\rightarrow \mathscr{B}}$. 
\end{definition}
In the quantum information theory or cryptographic context it is common to distinguish between classical and quantum inputs. For us the distinction is not needed, so we include any classical inputs into the $A_i$. We also note that ``Alice'' should be considered as an agency, which may have many agents Alice$_1$, Alice$_2$,... distributed throughout spacetime and co-operating according to pre-distributed instructions and subsequent communications. Finally, note that we make the convenient idealization that quantum systems can be localized to a spacetime point, though this is not strictly true due to holographic entropy bounds \cite{bousso2002holographic}. 

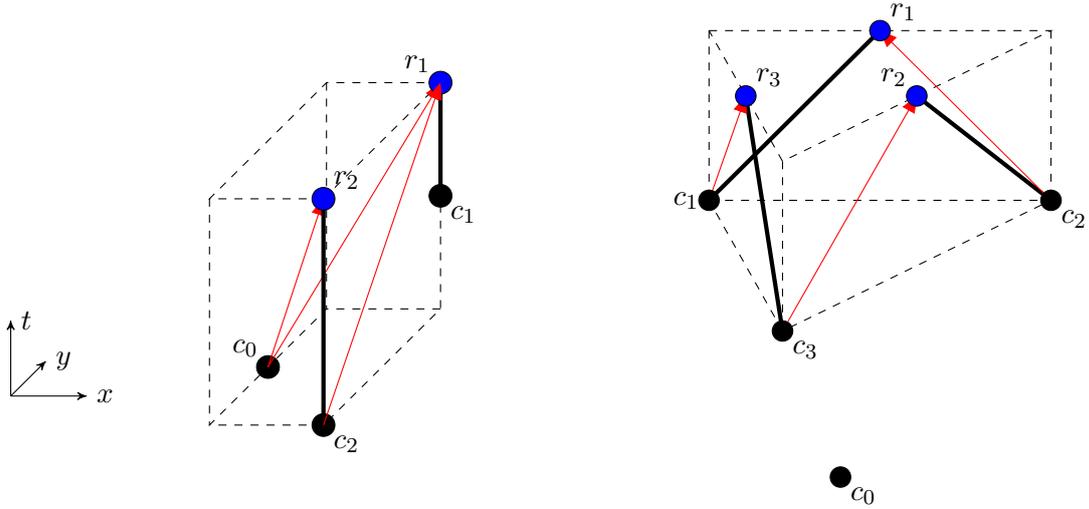
\begin{figure}
\centering
\begin{subfigure}{.45\textwidth}
\centering
\begin{tikzpicture}[scale=1]

	\coordinate (PT) at (0,0,0);
	
	\coordinate (CL) at (1.5,1.5,-2);
	\coordinate (CR) at (1.5,0,2);
	
	\coordinate (QL) at (1.5,3,-2);
	\coordinate (QR) at (1.5,3,2);
	
	\draw[dashed] (1.5,3,2) -- (1.5,3,-2) -- (1.5,0,-2) -- (1.5,0,2);
	\draw[dashed] (1.5,0,-2) -- (0,0,-2) -- (0,0,2) -- (1.5,0,2);
	\draw[dashed] (1.5,3,-2) -- (0,3,-2) -- (0,3,2) -- (1.5,3,2);
	\draw[dashed] (0,3,-2) -- (0,0,-2);
	\draw[dashed] (0,3,2) -- (0,0,2);
	
	\draw[fill=black] (PT) circle (0.15);
	\node [above left] at (PT) {$c_0$};
	
	\draw[fill=black] (CL) circle (0.15);
	\draw[fill=black] (CR) circle (0.15);
	
	\node [below right] at (CL) {$c_1$};
	\node [below right] at (CR) {$c_2$};
	
	\node [above left] at (QL) {$r_1$};
	\node [above right] at (QR) {$r_2$};
    
    \begin{scope}[shift={($(0,0)$)}]       
    \draw[->] (-3,0,1) -- (-3,0,-0.2);
    \node [right] at (-3,0,-0.2) {$y$};
    \draw[->] (-3,0,1) -- (-2,0,1);
    \node [right] at (-2,0,1) {$x$};
    \draw[->] (-3,0,1) -- (-3,1,1);
    \node [right] at (-3,1,1) {$t$};
    \end{scope}
    
	\draw[red][-triangle 45] (PT) -- (QL);
	\draw[red][-triangle 45] (PT) -- (QR);
    
	\draw[ultra thick] (CL) -- (QL);
	\draw[ultra thick] (CR) -- (QR);
    
	\draw[fill=blue]  (QL) circle (0.15);
	\draw[fill=blue]  (QR) circle (0.15);
    
    \draw[red][-triangle 45] (CR) -- (QL);
    
\end{tikzpicture}

\end{subfigure}
\hfill
\begin{subfigure}{.45\textwidth}
  \centering
  \begin{tikzpicture}[scale=0.9]

\coordinate (PS) at (2.5,-3.5,1.5);

	\coordinate (Ca) at (0,0,0);
	\coordinate (Cb) at (5,0,0);
	\coordinate (Cc) at (3,0,5);
	
	\coordinate (Qa) at (2.5,2.5,0);
	\coordinate (Qb) at (4, 2.5, 2.5);
	\coordinate (Qc) at (1.5, 2.5, 2.5);
	
	\draw[dashed] (Ca) -- (Cb) -- (Cc) -- (Ca);
	\draw[dashed] (0,2.5,0) -- (5,2.5,0) -- (3,2.5,5) -- (0,2.5,0);
	\draw[dashed] (0,2.5,0) -- (0,0,0);
	\draw[dashed] (5,2.5,0) -- (5,0,0);
	\draw[dashed] (3,2.5,5) -- (3,0,5);
	
	\draw[ultra thick] (Ca) -- (Qa);
	\draw[ultra thick] (Cb) -- (Qb);
	\draw[ultra thick] (Cc) -- (Qc);
	
	\draw[-triangle 45][red] (Ca) -- (Qc);
	\draw[-triangle 45][red] (Cb) -- (Qa);
	\draw[-triangle 45][red] (Cc) -- (Qb);
	
	\draw[fill=blue] (Qa) circle (0.15);
	\draw[fill=blue]  (Qb) circle (0.15);
	\draw[fill=blue]  (Qc) circle (0.15);
	
	\node [above right] at (Qa) {$r_1$};
	\node [above left] at (Qb) {$r_2$};
	\node [above right] at (Qc) {$r_3$};
	
	\draw[fill=black] (Ca) circle (0.15);
	\draw[fill=black] (Cb) circle (0.15);
	\draw[fill=black] (Cc) circle (0.15);
	
	\node [left] at (Ca) {$c_1$};
	\node [below right] at (Cb) {$c_2$};
	\node [below right] at (Cc) {$c_3$};
	
	\draw[fill=black] (PS) circle (0.15);
	\node [below right] at (PS) {$c_0$};

\end{tikzpicture}
\end{subfigure}
\caption{Two examples of quantum tasks. Red arrows indicate causal connections. In both tasks, Alice receives an unknown quantum state $\ket{\psi}$ at $c_0$ and classical bits $b_i$ at the $c_i$ for $i>0$. Exactly one of the $b_i$ will be $1$, which we label by $b_{i^*}$, while the others will be $0$. Alice doesn't know which will be $1$ in advance. She is required to return $\ket{\psi}$ at $r_{i^*}$. These are known as \emph{summoning tasks} \cite{kent2013no,hayden2016summoning,adlam2016quantum,kent2018unconstrained,hayden2018localizing}. Known protocols to complete the task at left use teleportation, while the known protocol for the task at right use an error correcting code with three shares that corrects for one erasure error. Figures reproduced from \cite{hayden2016summoning}.} 
\label{fig:summoning}
\end{figure}

There are restrictions on the class of relativistic quantum tasks that are possible, along with a set of tools for completing them. The obvious restrictions are no-cloning and no-signaling, but there are also more subtle restrictions \cite{kent2012quantum}. Other tasks are possible but require non-trivial strategies to complete, for instance teleportation or quantum error correction. A well studied subclass of such tasks are the generalized summoning tasks \cite{kent2013no,hayden2016summoning,adlam2016quantum,kent2018unconstrained,hayden2018localizing}, two simple examples of which are shown as figure \ref{fig:summoning}. 

Returning to the context of holography, we would like to phrase our operational holographic principle --- that what it is possible to accomplish in the bulk should be possible to accomplish in the boundary --- in more precise terms using the language of relativistic quantum tasks. Beginning with a bulk task it is not possible in general to identify a corresponding boundary task unambiguously, since a priori we do not have a boundary point or region that corresponds to a bulk input or output point. Starting with a boundary task it is straightforward to identify a corresponding bulk task however by simply embedding the boundary coordinates into the bulk spacetime. A bulk task with input and output points that may be identified with boundary points in this way we call an \emph{asymptotic quantum task} (AQT). 
To make this more concrete, consider global AdS$_{2+1}$. A suitable metric is given by
\begin{align}\label{eq:metric}
    ds^2 = -\cosh^2\rho\, dt^2 + d\rho^2 + \sinh^2 \rho \,d\varphi^2.
\end{align}
An asymptotic quantum task has inputs and outputs specified in the conformal boundary ($\rho=\infty$), which itself has metric
\begin{align}
    ds^2 = -dt^2 + d\varphi^2.
\end{align}
Thus, an asymptotic quantum task in AdS$_{2+1}$ is specified by
\begin{align}
    \mathbf{T} = \{\mathcal{M},\ket{\Psi}, \mathscr{A},\mathscr{B},\mathrm{c},\mathrm{r},\mathcal{N}\},
\end{align}
with 
\begin{align}
    \mathrm{c}&=\{c_i=(t_i,\varphi_i,\rho=\infty)\}, \nonumber \\
    \mathrm{r}&=\{r_i=(t_i,\varphi_i,\rho=\infty)\}.
\end{align}
We can identify this with the boundary task
\begin{align}
    \mathbf{\hat{T}} = \{\partial \mathcal{M},\ket{\psi},\mathscr{A},\mathscr{B},\hat{\mathrm{c}},\hat{\mathrm{r}},\mathcal{N}\}.
\end{align}
Some elements of the tuple that describes the bulk task have been modified to define the corresponding boundary task. In particular the output point are now
\begin{align}
    \hat{\mathrm{c}} &=\{\hat{c}_i=(t_i,\varphi_i)\} \nonumber \\
    \hat{\mathrm{r}}, &=\{\hat{r}_i=(t_i,\varphi_i)\}.
\end{align}
Further, the bulk geometry $\mathcal{M}$ has been replaced by its boundary $\partial \mathcal{M}$. Additionally, the state of the bulk fields $\ket{\Psi}$ is replaced by a corresponding state of the CFT, $\ket{\psi}$. 

Given this notion of an asymptotic quantum task and their corresponding boundary tasks, we can phrase our operational statement of the holographic principle as follows:
\begin{principle}\label{thm:AQT}
An asymptotic quantum task $\mathbf{T}$ is possible in the bulk if and only if the corresponding boundary task $\hat{\mathbf{T}}$ is.
\end{principle}
We will focus on the case where the bulk is described by low energy effective field theory. Because of this, we will typically only use the principle of asymptotic quantum tasks in one direction: given an asymptotic quantum task that can be completed in the bulk using the low energy dynamics, we use our principle to assert that there must be a boundary procedure for completing the same task. 

Given the restriction to using the implication of principle \ref{thm:AQT} only in one direction, from possible in the bulk to possible in the boundary, we have two ways in which the principle can be employed. First, if we find a task $\mathbf{T}$ and a bulk configuration $\ket{\Psi},\mathcal{M}$ such that $\hat{\mathbf{T}}$ is possible in the bulk and provably impossible in the boundary, then we can conclude that $\ket{\Psi}$, $\mathcal{M}$ must not correspond to a state occurring within the AdS/CFT correspondence\footnote{This is similar to recent work deriving energy conditions that must be obeyed by the bulk fields by beginning with boundary constraints \cite{lashkari2015inviolable,lashkari2016gravitational}. Field configurations that violate such energy conditions cannot occur within the correspondence.}. Alternatively, we can assume a particular bulk state is a valid state in the correspondence. Then, any task which is possible in that state must be possible in the boundary. In some cases this will imply constraints on the boundary state. 

Although usually we will have in mind a setting where the bulk is described as a classical spacetime on which various quantum fields may live, it is also possible to consider a more general scenario. For example, we could imagine our bulk region contains a black hole. Then the bulk would not be well described by a classical geometry near the singularity, but instead would require a quantum gravity based description. Nonetheless we can specify asymptotic quantum tasks in this background, since the input and output points occur in a region of this spacetime which is well described by classical geometry. Similarly, in section \ref{sec:procedures} we consider AdS/CFT with finite $N\gg 1$, so that the complete bulk description involves stringy corrections, but we may still discuss asymptotic quantum tasks. 

\begin{figure}
    \centering
    \begin{subfigure}{0.45\textwidth}
    \begin{tikzpicture}[scale=0.7]
    
    
    \node[below left] at (-4,0) {$c_1$};
    \draw[fill=black] (-4,0) circle (0.15);

    \node[below right] at (4,0) {$c_2$};
    \draw[fill=black] (4,0) circle (0.15);

    \node[below right] at (4,8) {$r_2$};
    \draw[fill=blue] (4,8) circle (0.15);

    \node[below left] at (-4,8) {$r_1$};
    \draw[fill=blue] (-4,8) circle (0.15);

    
    \draw[fill=gray,opacity=0.5] (-1,1) -- (1,1) -- (1,7) -- (-1,7) -- (-1,1);
    
    \end{tikzpicture}
    \caption{}
    \label{fig:taggingsub1}
    \end{subfigure}
    \hfill
\begin{subfigure}{.45\textwidth}
\begin{tikzpicture}[scale=0.7]

    \node[below left] at (-4,0) {$c_1$};
    \draw[fill=black] (-4,0) circle (0.15);

    \node[below right] at (4,0) {$c_2$};
    \draw[fill=black] (4,0) circle (0.15);

    \node[below right] at (4,8) {$r_2$};
    \draw[fill=blue] (4,8) circle (0.15);

    \node[below left] at (-4,8) {$r_1$};
    \draw[fill=blue] (-4,8) circle (0.15);
    
    \draw[postaction={on each segment={mid arrow}}] (-4,0) -- (-2,2) -- (-2,6) -- (-4,8);
    \draw[postaction={on each segment={mid arrow}}] (4,0) -- (2,2) -- (2,6) -- (4,8);
    \draw[postaction={on each segment={mid arrow}}] (-2,2) -- (0,4) -- (2,6);
    \draw[postaction={on each segment={mid arrow}}] (2,2) -- (0,4) -- (-2,6);
    
    \draw[dashed] (2,2) -- (0,0) -- (-2,2);
    \node[below] at (0,0) {$\ket{\Psi^+}^{\otimes n}$};
    
    \draw[fill=yellow] (-2,2) circle (0.3);
    \draw[fill=yellow] (2,2) circle (0.3);
    \draw[fill=yellow] (-2,6) circle (0.3);
    \draw[fill=yellow] (2,6) circle (0.3);
    
    \draw[fill=gray,opacity=0.5] (-1,1) -- (1,1) -- (1,7) -- (-1,7) -- (-1,1);
    
\end{tikzpicture}
\caption{}
\label{fig:taggingsub2}
\end{subfigure}
    
\caption{a) A quantum tagging scheme: Bob asks Alice to complete a quantum task which he hopes requires Alice perform quantum operations within a designated spacetime region (shown in grey). b) A ``spoof'' of Bob's tag. Alice performs operations at the four spacetime locations shown as yellow dots (which lie outside the grey region), between which she exchanges a round of communication (which may pass through the grey region). Results in non-local quantum computation show that this replacement is always possible \cite{vaidman2003instantaneous,buhrman2014position}. The dashed lines indicate entangled states have been shared between the two lower yellow dots. A concrete protocol for performing a spoof is given in section \ref{sec:teleportationspoof}.}
\label{fig:tagging}
\end{figure}
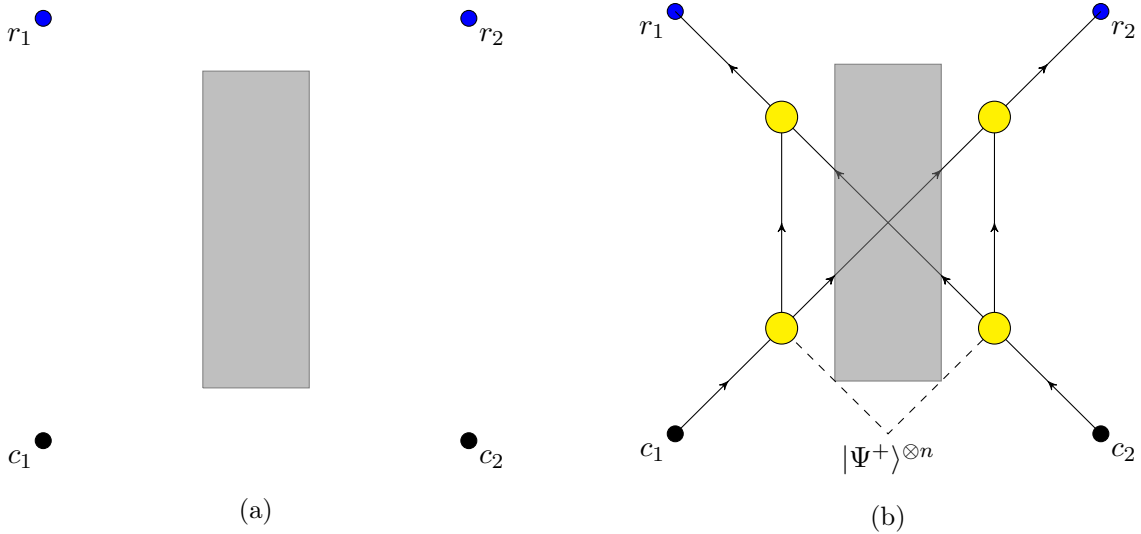

Historically the interest in relativistic quantum tasks has been partly due to cryptographic applications. Some notable successes in this program include bit commitment \cite{kent2012unconditionally,kent2011unconditionally}, coin flipping \cite{kent1999coin} and work on key distribution \cite{barrett2005no}. Another cryptographic goal that has been studied in the context of relativistic quantum tasks is position verification, also known as `quantum tagging'. In a quantum tagging scheme Bob's goal is to verify Alice's spatial location, without himself visiting that location. Consider for instance the arrangement of input and output points shown as figure \ref{fig:taggingsub1}. Bob will give inputs at the $c_i$ and expect certain outputs at the $r_i$. Bob's hope is that, by choosing carefully his inputs and the quantum channel he expects Alice to do, he can force Alice to perform that channel within a certain spacetime region. Unfortunately for Bob there is no channel he can choose that will force Alice to apply it within the designated spacetime region. Instead, it is always possible to replace an operation performed in the central spacetime region with operations performed outside the region, along with pre-shared entanglement and a single two way exchange of information \cite{vaidman2003instantaneous,buhrman2014position}. We illustrate this in figure \ref{fig:taggingsub2}. 

To incorporate quantum tagging into the framework of quantum tasks, Kent \cite{kent2012quantum} considered spacetime regions in which Alice may only perform a limited class of operations. In the context of quantum tagging in a Minkowski space background, the relevant restriction is to consider regions through which quantum or classical signals may be sent but within which no quantum operations may be performed. Another possibility is to exclude Alice entirely from a region, preventing even signals from being sent through. 

An asymptotic quantum task in AdS space can be viewed as a tagging scheme. To do this, we designate the bulk of AdS as an entirely excluded region in the stronger sense above, excluding even signals through the region. Spoofing schemes then become methods for completing asymptotic quantum tasks in the boundary. To transform a spoofing scheme into a boundary procedure, consider figure \ref{fig:taggingsub2}. The input and output points are now points in the boundary of AdS. The yellow dots, which signify regions where the spoof requires certain quantum operations be performed, are taken infinitesimally close to their nearby input or output points. The black lines between the yellow dots, which signify classical signals, are wrapped around the boundary of AdS. It is always possible to have the black lines remain as causal curves, because points that are connected through the bulk spacetime are always connected through the boundary \cite{gao2000theorems,engelhardt2016gravity}. The AdS/CFT dictionary also provides a method for completing asymptotic quantum tasks, since once a bulk procedure is provided the dictionary can be used to translate this into a boundary procedure. We call any method for completing asymptotic quantum tasks a \emph{holographic procedure}. Notice that while a dictionary provides a correspondence between a bulk procedure and a boundary procedure, a holographic procedure need only provide the boundary perspective. Further, this boundary procedure need not be tied to any bulk one. 

\section{Implications for holography from asymptotic quantum tasks}\label{sec:implications}

In this section we take as our guiding principle the implication from bulk to boundary tasks, which says that for tasks occurring within the AdS/CFT correspondence we have
\begin{align}\label{eq:AQTprinciple}
    \text{AQT possible in bulk} \implies \text{AQT possible in the boundary}.
\end{align}
Sections \ref{sec:causality} and \ref{sec:causalwedge} reproduce some known features of AdS/CFT. Section \ref{sec:entanglement-geometry} develops a new result in the relationship between entanglement and geometry.

\subsection{Bulk and boundary causality}\label{sec:causality}

\begin{figure}
\begin{center}
    \tdplotsetmaincoords{10}{0}
    \begin{tikzpicture}[scale=1,tdplot_main_coords]
    \tdplotsetrotatedcoords{0}{-50}{0}
    \draw (-2,0,0) -- (-2,4,0);
    \draw (2,0,0) -- (2,4,0);
    
    \begin{scope}[tdplot_rotated_coords]
    \begin{scope}[canvas is xz plane at y=0]
    \draw (0,0) circle [radius=2];
    \end{scope}
    
    \begin{scope}[canvas is xz plane at y=4]
    \draw (0,0) circle [radius=2];
    \end{scope}
    
    \draw[red,-triangle 45] (-2,0,0) -- (2,5,0);
    
    \draw plot [mark=*, mark size=2] coordinates{(-2,0,0)};
    \node[below left] at (-2,0,0) {$c$};
    \draw[->] (-2,-0.5,0) -- (-2,0,0);
    \node[below] at (-2,-0.5,0) {$\ket{\psi}$};
    
    \draw[blue] plot [mark=*, mark size=2] coordinates{(2,4,0)};
    \node[above right] at (2,4,0) {$r$};
    \draw[->] (2,4,0) -- (2,4.5,0);
    \node[above right] at (2,4.5,0) {$\ket{\psi}$};
    
    \draw[domain=0:180] plot ({-2*cos(\x)},{\x/45},{-2*sin(\x)});
    
    \end{scope}
    \end{tikzpicture} 
\end{center}
\caption{The quantum task discussed in section \ref{sec:causality}. A quantum state is received at a point on the boundary $c$ and must be returned at the point $r$. Implication \ref{eq:AQTprinciple} implies that signals cannot travel faster through the bulk than through the boundary, consistent with the usual statement of boundary causality \cite{gao2000theorems,engelhardt2016gravity}.}
\end{figure}
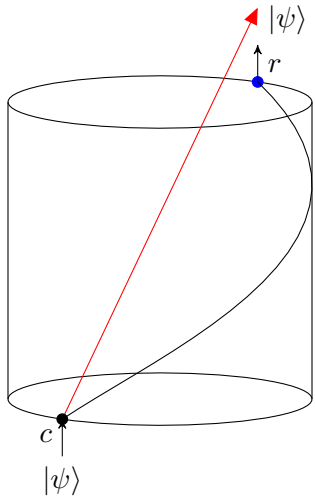

A simple but non-trivial quantum task consists of a single input point $c$ and single output point $r$. We specify that at $c$ Alice receives a quantum state $\ket{\psi}$ which she must return at $r$. In the bulk picture, Alice can complete this task whenever $c \prec r$. Since the task being possible in the bulk implies it is possible in the boundary, we get that $c \prec r$ in the bulk geometry implies $\hat{c} \prec \hat{r}$ in the boundary geometry. Stated differently, a signal cannot travel faster through the bulk than through the boundary. This is just the usual statement relating bulk and boundary causality \cite{gao2000theorems,engelhardt2016gravity}. 

\subsection{A property of causal wedges}\label{sec:causalwedge}

Among the best studied quantum tasks are the summoning tasks \cite{kent2013no,hayden2016summoning}, which we introduced in figure \ref{fig:summoning}, and their variants \cite{adlam2016quantum,hayden2018localizing}. A summoning task is defined by a start point $c_0$ and a set of call-reveal pairs $(c_i,r_i)$. At the start point Alice receives a quantum state $\ket{\psi}$. At the $c_i$ Bob outputs a bit $b_i\in \{0,1 \}$. Alice has a guarantee that exactly one of the $b_i$ will have $b_i=1$ and the remainder will have $b_i=0$. She is required to return $\ket{\psi}$ at $r_{i_*}$ such that $b_{i_*}=1$. 

To characterize summoning tasks it is helpful to consider the causal diamonds defined by $D_i = J^+(c_i)\cap J^-(r_i)$. These represent the spacetime region in which Alice both knows the call information from $c_i$ and can act on it if she needs to return the state to $r_i$. It is also useful to specify causal relations among the diamonds by saying that diamonds $D_1$ and $D_2$ are causally connected if there is a causal curve that passes from one diamond to the other. The following theorem characterizes when a summoning task is possible.
\begin{theorem}
A summoning task with two diamonds $D_1$, $D_2$ is possible if and only if the following two conditions are true.
\begin{enumerate}
    \item There is a causal curve from the start point $c_0$ through $D_1$ and a causal curve from the start point $c_0$ through $D_2$.
    \item $D_1$ and $D_2$ are causally connected.
\end{enumerate}
\end{theorem}
In the context of holography the summoning theorem translates into a simple property of the causal wedge. In particular, it gives that $\mathcal{C}(D_1)$ and $\mathcal{C}(D_2)$ are causally disconnected whenever $D_1$ and $D_2$ are. To see this, suppose we have $D_1, D_2$ which are causally disconnected, and consider a summoning task on $D_1$ and $D_2$ with $c_0$ at a sufficiently early time to be in their causal past. By the summoning theorem this task is impossible in the boundary, but then by our principle of asymptotic quantum tasks \ref{eq:AQTprinciple} it is impossible in the bulk. The bulk task shares the same call and reveal points, but now the relevant causal diamonds are in the bulk geometry and coincide with the causal wedges ${J}^+(c_i)\cap {J}^-(r_i) = \mathcal{C}(D_i)$. Since the bulk task is impossible, applying the no-summoning theorem again we have that $\mathcal{C}(D_1)$ and $\mathcal{C}(D_2)$ are causally disconnected, as needed. This property of the causal wedge has been noted before \cite{hubeny2013global}. 

\subsection{An entanglement-causal structure connection}\label{sec:entanglement-geometry}

In the introduction we mentioned the quantum task of quantum tagging \cite{buhrman2014position,kent2006tagging}. In this section we employ results on tagging to our framework of asymptotic quantum tasks. We will employ results on the necessity of entanglement in quantum tagging to reach conclusions about which boundary CFT regions must be entangled at $O(N^2)$. Doing so requires some extensions of the existing results on tagging, which we develop in the appendices.

\begin{figure}
    \centering
    \begin{subfigure}{.45\textwidth}
    \centering
    \tdplotsetmaincoords{10}{0}
    \begin{tikzpicture}[scale=1.0,tdplot_main_coords]
    \tdplotsetrotatedcoords{0}{20}{0}
    \draw (-2,0,0) -- (-2,4,0);
    \draw (2,0,0) -- (2,4,0);
    
    \begin{scope}[tdplot_rotated_coords]
    \begin{scope}[canvas is xz plane at y=0]
    \draw (0,0) circle [radius=2];
    \end{scope}
    
    \begin{scope}[canvas is xz plane at y=4]
    \draw (0,0) circle [radius=2];
    \end{scope}
    
    \draw plot [mark=*, mark size=2] coordinates{(2,0,0)};
    \node[right] at (2,0,0) {$c_2$};
    \draw[->] (2,-0.5,0) -- (2,0,0);
    \node[below] at (2,-0.5,0) {$q$};
    
    \draw plot [mark=*, mark size=2] coordinates{(-2,0,0)};
    \node[left] at (-2,0,0) {$c_1$};
    \draw[->] (-2,-0.5,0) -- (-2,0,0);
    \node[below] at (-2,-0.5,0) {$H^q\ket{b}$};
    
    \draw[blue] plot [mark=*, mark size=2] coordinates{(0,4,-2)};
    \node[below left] at (0,4,-2) {$r_1$};
    \draw[->] (0,4,-2) -- (0,4.5,-2);
    \node[above] at (0,4.5,-2) {$b$};
    
    \draw[blue] plot [mark=*, mark size=2] coordinates{(0,4,2)};
    \node[below right] at (0,4,2) {$r_2$};
    \draw[->] (0,4,2) -- (0,4.5,2);
    \node[above] at (0,4.5,2) {$b$};
    
    \end{scope}
    
    \end{tikzpicture} 
    \end{subfigure}
    \hfill
    \begin{subfigure}{.45\textwidth}
    \centering
    \begin{tikzpicture}[scale=1.25]
    
    \draw (-2,0) -- (2,0) -- (2,2) -- (-2,2) -- (-2,0);
    
    \draw[black] plot [mark=*, mark size=2] coordinates{(-2,0)};
    \draw[->] (-2,-0.5)--(-2,-0.05);
    \node[below] at (-2,-0.5) {$q$};
    \node[below left] at (-2,0) {$c_2$};
    
    \draw[black] plot [mark=*, mark size=2] coordinates{(2,0)};
    \node[below right] at (2,0) {$c_2$};
    
    \draw[black] plot [mark=*, mark size=2] coordinates{(0,0)};
    \draw[->] (0,-0.5)--(0,-0.05);
    \node[below] at (0,-0.5) {$H^q\ket{b}$};
    \node[below left] at (0,0) {$c_1$};
    
    \draw[->] (-1,2) -- (-1,2.5);
    \node[above] at (-1,2.5) {$b$};
    \draw[blue] plot [mark=*, mark size=2] coordinates{(-1,2)};
    \node[above right] at (-1,2) {$r_1$};
    
    \draw[->] (1,2) -- (1,2.5);
    \node[above] at (1,2.5) {$b$};
    \draw[blue] plot [mark=*, mark size=2] coordinates{(1,2)};
    \node[above right] at (1,2) {$r_2$};
    
    \node at (0,-1.5) {$ $};
    \node at (0,3.5) {$ $};
    
    \end{tikzpicture}
    \end{subfigure}
    \caption{Bulk and boundary perspectives on the $\mathbf{{B}}_{84}$ task. Generically, it may happen that the bulk geometry contains a central region $P=J^+(c_1)\cap J^+(c_2)\cap J^-(r_1)\cap J^-(r_2)$ while the boundary geometry does not. In order for the boundary task to be possible when the bulk one is, the boundary must make up for the lack of central region by having $O(N^2)$ entanglement between certain pairs of boundary regions, as we argue for in detail in the text.}
    \label{fig:bulkandboundarybb84}
    \end{figure}
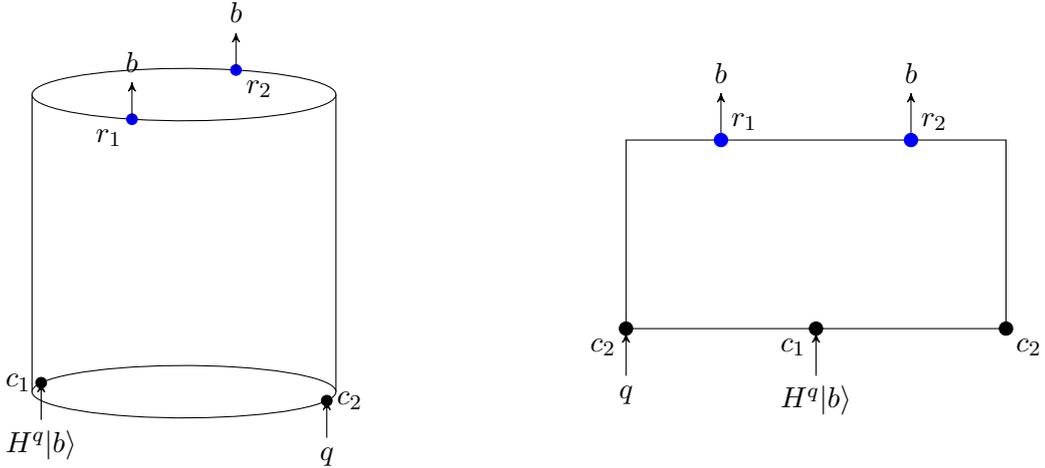

We begin by constructing an asymptotic quantum task, which we call $\mathbf{B}_{84}^{\times n}$. The basic task $\mathbf{B}_{84}^{\times 1}$ has two input points $c_1$, $c_2$ and two output points $r_1$, $r_2$. The inputs $\mathscr{A}$ and outputs $\mathscr{B}$ are
\begin{align}
    A_1 &= H^q\ket{b} \,\,\,\,\,\,  B_1 = \ket{b} \nonumber \\
    A_2 &= \ket{q}\,\,\,\,\,\,\,\,\,\,\,\,\, B_2 = \ket{b}.
\end{align}
To form $\mathbf{B}_{84}^{\times n}$ we repeat $\mathbf{B}_{84}^{\times n}$ in parallel $n$ times. The inputs and outputs are now,
\begin{align}
    A_1 &= H^{q_1}\ket{b_1}H^{q_2}\ket{b_2}...H^{q_n}\ket{b_n} \,\,\,\,\,\,\,\,\,\,\,  B_1 = \ket{b_1}\ket{b_2}...\ket{b_n} \nonumber \\
    A_2 &= \ket{q_1}\ket{q_2}...\ket{q_n}\,\,\,\,\,\,\,\,\,\,\,\,\,\,\,\,\,\,\,\,\,\,\,\,\,\,\,\,\,\,\,\,\,\,\,\,\,\,\,\,\, B_2 = \ket{b_1}\ket{b_2}...\ket{b_n}.
\end{align}
To visualize the task and in the context of an explicit calculation we perform later, it is useful to consider this task in AdS$_{2+1}$. The $\mathbf{B}_{84}^{\times 1}$ task in AdS$_{2+1}$ is illustrated in figure \ref{fig:bulkandboundarybb84}. 

The most straightforward procedure by which Alice can complete the $\mathbf{{B}}_{84}^{\times n}$ task in the bulk perspective is as follows. Alice brings each of her inputs $H^{q_i}\ket{b_i}$ and $q_i$ together in a region $P$, applies $H^{q_i}$ to $H^{q_i}\ket{b_i}$ to get $\ket{b_i}$, then copies the $\ket{b_i}$ in the computational basis and sends a copy to each of $r_1$ and $r_2$. This procedure can be followed whenever the \emph{central region} $P=J^+(c_1)\cap J^+(c_2) \cap J^-(r_1)\cap J^-(r_2)$ is non-empty, and $n$ is small enough so that the qubits sent through this region do not back-react significantly. The most interesting case occurs when the same intersection of light cones considered in the boundary theory is empty, that is $\hat{P}=\emptyset$. Then, while this simple procedure cannot be used in the boundary, our principle of asymptotic quantum tasks implies that there must be some other procedure to complete the task in the boundary.

To see that it can occur that $\hat{P}=\emptyset$ and $P\neq \emptyset$, consider the following choice of locations for the input and output points of a $\mathbf{B}_{84}^{\times n}$ task defined in AdS$_{2+1}$: 
\begin{align}\label{eq:ads3example}
    c_1 &= (-x/2,-\alpha-x/2,\infty), \,\,\,\,\,\,\,\,\,\,\,\,\,\,\,\,\,\,\, r_1 = (\alpha+x,0,\infty), \nonumber \\
    c_2 &= (-x/2,\alpha+x/2,\infty), \,\,\,\,\,\,\,\,\,\,\,\,\,\,\,\,\,\,\,\,\,\,\, r_2 = (\pi-\alpha,\pi,\infty).
\end{align}
In this setting we have that $\hat{P}=\emptyset$ so long as $\alpha+x<\pi$. As we show in appendix \ref{app:minsurfaceandbulkpoint}, for pure AdS$_{2+1}$ the bulk central region is non-empty whenever
\begin{align}
    \sin^2(x/2) \geq \sin(x+\alpha)\sin(\alpha).
\end{align}
Consequently there will be many points in the $(\alpha,x)$ parameter space where $\hat{P}=\emptyset$ and $P \neq \emptyset$. One example occurs with $x=\pi/2,\alpha=\pi/4$.

\begin{figure}
    \centering
    \begin{tikzpicture}[scale=0.75]
    
    \draw[thick, black, fill=black!60!,opacity=0.8] (-6,0) -- (-4,2) -- (-2,0) -- (-4,-2) -- (-6,0);
    
    \draw[thick, black, fill=black!60!,opacity=0.8] (6,0) -- (4,2) -- (2,0) -- (4,-2) -- (6,0);
    
    \draw[lightgray] (-10,8) -- (10,8) -- (10,-2) -- (-10,-2) -- (-10,8);
    
    \draw[dashed] (-6,0) -- (0,6);
    \draw[dashed] (6,0) -- (0,6);
    \draw[dashed] (-2,0) -- (-10,8);
    \draw[dashed] (2,0) -- (10,8);

    \draw[->] (0,6) --(0,7);
    \draw[->] (10,8) -- (10,9);
    \draw[->] (-10,8) -- (-10,9);
    
    \node[left] at (-0,7) {$b$};
    \node[left] at (10,9) {$b$};
    \node[right] at (-10,9) {$b$};
    
    \draw[fill=black] (-4,-2) circle (0.15);
    \draw[fill=black] (4,-2) circle (0.15);
    \node[below left] at (-4,-2) {$c_1$};
    \node[below right] at (4,-2) {$c_2$};
    
    \draw[->] (-4,-3) -- (-4,-2.18);
    \draw[->] (4,-3) -- (4,-2.18);
    \node[right] at (-4,-3) {$H^q\ket{b}$};
    \node[left] at (4,-3) {$q$};
    
    \draw[->] (6.75,-1.5) to  [out=180,in=-50] (4.25,0);
    \node[right] at (6.75,-1.5) {$R_2$};
    
    \draw[->] (-6.75,-1.5) to  [out=0,in=-130] (-4.25,0);
    \node[left] at (-6.75,-1.5) {$R_1$};
    
    \draw[ultra thick,green] (-6,0) -- (-2,0);
    \draw[ultra thick,green] (2,0) -- (6,0);
    
    \draw[fill=blue] (10,8) circle (0.15);
    \draw[fill=blue] (-10,8) circle (0.15);
    \draw[fill=blue] (0,6) circle (0.15);
    \node[above left] at (-10,8) {$r_2$};
    \node[above right] at (10,8) {$r_2$};
    \node[above right] at (0,6) {$r_1$}; 
    
    \end{tikzpicture}
    \caption{The $\mathbf{\hat{B}}_{84}^{\times 1}$ task. The task takes place in $R\times S^1$, with the right and left vertical grey lines identified to form the $S^1$. At $c_1$ Alice receives the state $H^q \ket{b}$ with $b\in\{0,1\}$ and at $c_2$ she receives the classical bit $q$. At $r_1$ and $r_2$ Alice is required to return $b$. The grey diamonds are the input regions, defined by $D(R_i) = J^+(c_i)\cap J^-(r_1) \cap J^-(r_2)$.}
    \label{fig:BB84task}
\end{figure}
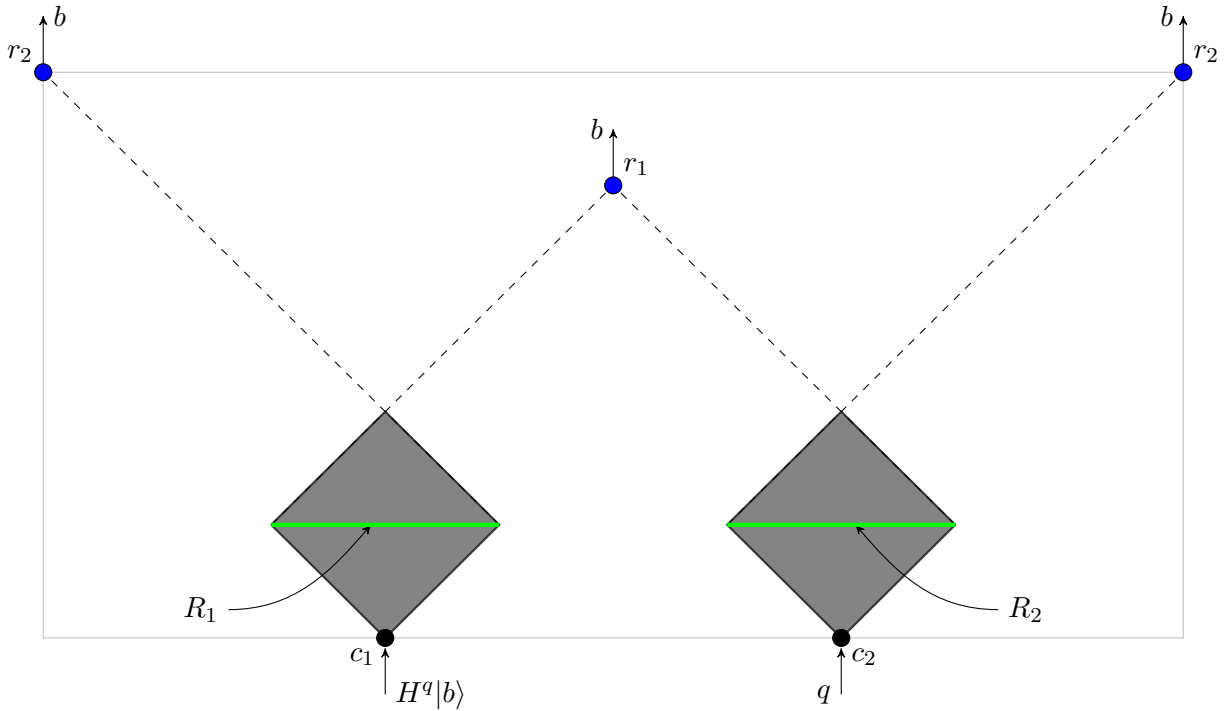

It has been shown that whenever the boundary central region is empty every method for completing $\mathbf{\hat{B}}_{84}$ must make use of entanglement shared between near $c_1$ and near $c_2$. More precisely, we will argue that the entanglement should be between the \emph{input regions} $R_1$ and $R_2$, defined by
\begin{align}
    D(R_i) = \hat{J}^+(c_i) \cap \hat{J}^-(r_1)\cap \hat{J}^-(r_2).
\end{align}
Intuitively, this is the region in which a boundary observer can apply quantum operations to systems provided at $c_i$ and send her outputs to either of $r_1$ or $r_2$. The regions $R_1$ and $R_2$ are illustrated in figure \ref{fig:BB84task}.

We give two results on the necessity of entanglement in the $\textbf{B}_{84}^{\times n}$ task below. In the proofs of these theorems we consider only the state on the regions $R_1$ and $R_2$ as defined above. In fact though, there are additional spacetime regions between the $R_i$ which are in the past of the output points $r_1, r_2$. Call these regions $X_1$ and $X_2$, with $X_1$ in the past of $r_1$ and $X_2$ in the past of $r_2$. It is possible that the additional degrees of freedom living in the $X_i$ may make completing the $\textbf{B}^{\times n}_{84}$ task possible even without entanglement between the $R_i$. We will assume throughout however that this is not the case, and treat only the (mixed) state $\rho_{R_1R_2}$. For this reason the conclusion of this section will be stated as a conjecture and not a theorem.\footnote{The main theorem of this section has since been proven using classical general relativity in \cite{may2019holographic}. They also give an improved version of our argument, but it is not yet a proof.}

We give the first result on necessity of entanglement below, which is taken from \cite{tomamichel2013monogamy}. 
\begin{theorem}\label{thm:necessityofentanglement}
Consider the $\mathbf{\hat{B}}_{84}^{\times n}$ task, with input points $c_1,c_2$ and output points $r_1,r_2$. Then if the central region $\hat{P}$ is empty and the resource state $\rho_{R_1R_2}$ has $I(R_1:R_2)=0$, then the task can be completed with probability at most $\beta^n$ where $\beta = \frac{1}{2}+ \frac{1}{2\sqrt{2}}\approx 0.85$.
\end{theorem}
We recall the proof in appendix \ref{app:necessityproof}. Applying the principle of asymptotic quantum tasks to this theorem, we find that whenever the bulk central region is non-empty we must have that $R_1$ and $R_2$ have positive mutual information. This is a weak result however, as in a strongly interacting CFT we generically expect all subregions to have $O(N^0)$ mutual information.

We can arrive at a stronger conclusion by considering more precisely the boundary resources required to complete the $\mathbf{\hat{B}}_{84}^{\times n}$ task, which we quantify in the following theorem.
\begin{theorem}\label{thm:linearbound}
For any state $\rho_{R_1R_2}$ which suffices to complete the $\mathbf{B}_{84}^{\times n}$ task with success probability $1$, the mutual information $I(R_1:R_2)$ is bounded below according to
\begin{align}
\frac{1}{2}I(R_1:R_2) \geq  n \left(\log_2 1/\beta \right) - 1,
\end{align}
where $\beta = \frac{1}{2}+ \frac{1}{2\sqrt{2}}\approx 0.85$ so that $\log_21/\beta \approx 0.23$. 
\end{theorem}
We prove this theorem in appendix \ref{app:linearbound}. Notice that for $\mathbf{\hat{B}}_{84}^{\times n}$ a linear bound is the tightest possible, since protocols are known which allow the task to be completed with perfect success probability using $n$ EPR pairs. Such a protocol is also given in appendix \ref{app:linearbound}. 

Given a $\mathbf{B}_{84}^{\times n}$ task with a particular choice of input and output points, we can ask if the task is possible in the bulk perspective. If the bulk central region $P=J^-(r_1)\cap J^-(r_2) \cap J^+(c_1)\cap J^+(c_2)$ is non-empty and $n$ is small enough to avoid back-reacting on the geometry and closing $P$, then we can conclude the task is possible in the bulk. To understand when back reaction will become significant, we consider that the $n$ qubits entering the region $P$ will be associated with some energy content and this energy will couple to the metric with a factor of Newtons constant $G$. Thus the strength of the coupling is proportional to $G (\Delta E) n$ where $\Delta E$ is the energy carried by each qubit. In AdS/CFT Newtons constant scales like $N^{-2}$, for $N$ the size of the gauge symmetry in the boundary CFT. Consequently, at large $N$ the $n$ qubits will have negligible back reaction so long as $n < O(N^2)$. We can conclude then that $\mathbf{B}^{\times n}_{84}$ is possible whenever the central region $P$ is non-empty and $n<O(N^2)$.

We can now apply the principle of asymptotic quantum tasks and conclude that, in the boundary, $\mathbf{\hat{B}}_{84}^{\times n}$ is possible whenever the bulk central region $P$ is non-empty and $n < O(N^2)$. From theorem \ref{thm:linearbound}, and now making our assumption that state on the additional spacetime regions $X_1$ and $X_2$ plays no role, we then have that the boundary in-regions $R_1$ and $R_2$ share a mutual information bounded below by a constant times $n$. Choosing $n$ to be larger than $O(N^0)$ while still less than $O(N^2)$ to avoid back reaction, we get that $I(R_1:R_2)$ is larger than $O(1)$. In fact, since in a holographic CFT the mutual information decomposes into an $O(N^0)$ term and a $O(N^2)$ term, we can conclude that the mutual information is of order $O(N^2)$. We summarize this statement as the following conjecture.
\begin{conjecture}\label{conj:entangledregions}
Consider four spacetime points $c_1,c_2,r_1$ and $r_2$. Then if the bulk central region is non-empty and the boundary central region is empty, then we have that $I(R_1:R_2)$ is $O(N^2)$.
\end{conjecture}
The argument for the conjecture applies in any dimension. Recall also that the central region is defined by $P=J^-(r_1)\cap J^-(r_2) \cap J^+(c_1)\cap J^+(c_2)$, the boundary central region is the same intersection of light cones but taken in the boundary geometry, and the regions $R_i$ are defined according to $D(R_i) = J^+(c_i) \cap J^-(r_1)\cap J^-(r_2)$.

Notice that the implication in conjecture \ref{conj:entangledregions} is stated in only one direction, due in part to the caveat stressed earlier that boundary procedures may translate to bulk procedures which go beyond the bulk effective field theory. The reverse implication is also prevented however due to the entanglement measure we have used. The mutual information being positive does not imply the $\mathbf{\hat{B}}_{84}^{\times n}$ task can be completed, since the mutual information counts classical as well as quantum correlations, and classical correlations are insufficient to complete $\mathbf{\hat{B}}_{84}^{\times n}$.

\begin{figure}
    \centering
    \begin{subfigure}{0.45\textwidth}
    \begin{tikzpicture}
    
    \draw[thick] (0,0) circle (3);
    
    \draw [domain=60:120,fill=lightgray,opacity=0.8] plot ({3*cos(\x)}, {3*sin(\x)}) -- (-1.5, 2.60) to [out=-60,in=-120] (1.5, 2.60);
     \draw [domain=-60:-120,fill=lightgray,opacity=0.8] plot ({3*cos(\x)}, {3*sin(\x)}) -- (-1.5, -2.60) to [out=60,in=120] (1.5, -2.60);
    
    \draw [green,ultra thick,domain=60:120] plot ({3*cos(\x)}, {3*sin(\x)});
    \draw [green,ultra thick,domain=-60:-120] plot ({3*cos(\x)}, {3*sin(\x)});
    
    \draw[blue, thick] (1.5, 2.60) to [out=-120,in=-60] (-1.5, 2.60);
    \draw[blue, thick] (1.5, -2.60) to [out=120,in=60] (-1.5, -2.60);
    
    \draw[<->,gray,domain=59:-59] plot ({3.3*cos(\x)},{3.3*sin(\x)});
    \node[right,gray] at (3.3,0) {$2\alpha$};
    
    \draw[<->,gray,domain=61:119] plot ({3.3*cos(\x)},{3.3*sin(\x)});
    \node[above,gray] at (0,3.3) {$x$};
    
    \end{tikzpicture}
    \caption{}
    \label{fig:disconnectedsurfaces}
    \end{subfigure}
    \hfill
    \begin{subfigure}{0.45\textwidth}
    \begin{tikzpicture}
    
    \draw [lightgray,domain=80:20,fill=lightgray,opacity=0.8] plot ({3*cos(\x)}, {3*sin(\x)}) -- (2.82,1.03) to [out=-160,in=+90] (2.25,0) -- (0.205,0) to [out=90,in=-100] (0.52,2.96);
    \draw [lightgray,domain=-80:-20,fill=lightgray,opacity=0.8] plot ({3*cos(\x)}, {3*sin(\x)}) -- (2.82,-1.03) to [out=160,in=-90] (2.25,-0.012) -- (0.205,-0.012) to [out=-90,in=100] (0.52,-2.96);   
    
    \draw[thick] (0,0) circle (3);
    \draw [green,ultra thick,domain=20:80] plot ({3*cos(\x)}, {3*sin(\x)});
    \draw [green,ultra thick,domain=-20:-80] plot ({3*cos(\x)}, {3*sin(\x)});
    
    \draw[blue,thick] (2.82,1.03) to [out=-160,in=+160] (2.82,-1.03);
    \draw[blue,thick] (0.52,2.96) to [out=-100,in=100] (0.52,-2.96);
    
    \draw[<->,gray,domain=19:-19] plot ({3.3*cos(\x)},{3.3*sin(\x)});
    \node[right,gray] at (3.3,0) {$2\alpha$};
    
    \draw[<->,gray,domain=21:79] plot ({3.3*cos(\x)},{3.3*sin(\x)});
    \node[above right,gray] at (2.12,2.52) {$x$};
    
    \node[above,gray] at (0,3.3) {$ $};
    
    \end{tikzpicture}
    \caption{}
    \label{fig:connectedsurfaces}
    \end{subfigure}
    \caption{Minimal surfaces (shown in blue) for two intervals $R_1$ and $R_2$ (shown in green) of equal size $x$ sitting on a constant time slice of AdS$_{2+1}$. The intervals are separated by an angle $2\alpha$. The entanglement wedge $\mathcal{E}(R_1R_2)$ (shown in grey) is the region whose boundary is the union of the regions $R_1$ and $R_2$ and their minimal surfaces. For large $\alpha$ and small $x$ the entanglement wedge of the region $R_1\cup R_2$ is disconnected, while for small $\alpha$ or large  enough $x$, the entanglement wedge becomes connected, as shown at right. The entanglement wedge being connected indicates the mutual information is $O(N^2)$, while a disconnected entanglement wedge indicates the mutual information is $O(N^0)$.}
    \label{fig:min_surfaces}
\end{figure}
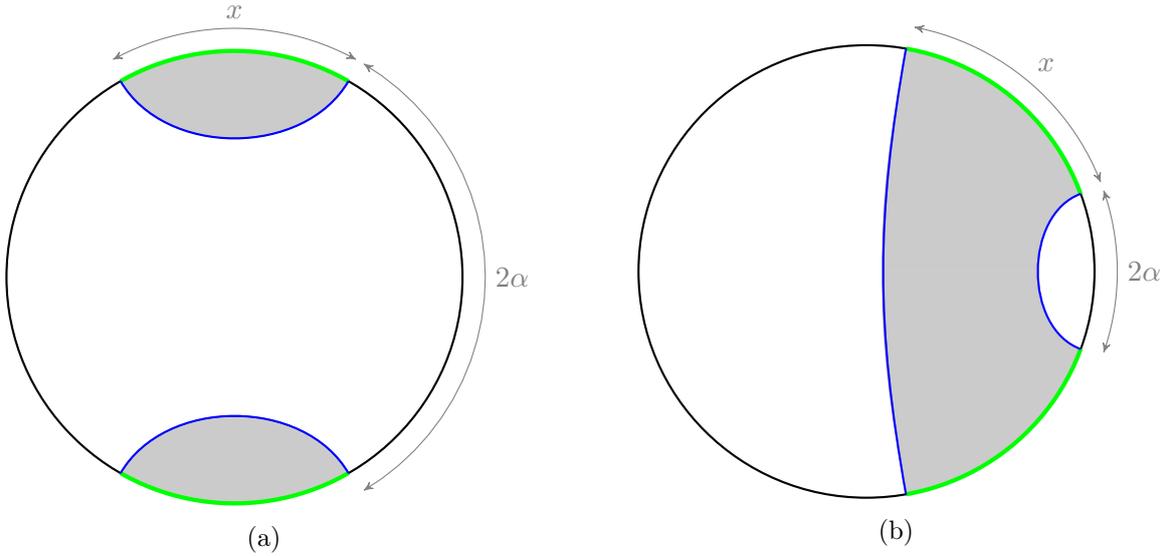

In AdS/CFT we have an independent method for understanding when two boundary regions are entangled at order $O(N^2)$. The Ryu-Takayanagi (RT) formula relates minimal surfaces in the bulk to boundary entanglement entropy according to
\begin{align}\label{eq:RTformula}
    S(R) = \frac{\mathcal{A}(R)}{4G} + S(\mathcal{E}(R)),
\end{align}
where $\mathcal{A}(R)$ denotes the area of the minimal bulk surface ending on $R$ and $S(\mathcal{E}(R))$ is the entropy of any quantum fields present in the entanglement wedge of $R$. The area term is $O(N^2)$ while the bulk entanglement term is $O(1)$. The mutual information is written in terms of the entanglement entropy according to
\begin{align}
    I(R_1:R_2)\equiv S(R_1) + S(R_2) - S(R_1R_2).
\end{align}
Inserting the RT formula into this expression, we get that the mutual information is given by an order $O(N^2)$ term calculated using the areas of minimal surfaces, and an $O(1)$ term given by the bulk entropy terms,
\begin{align}\label{eq:mutualinfo}
    I(R_1:R_2) = &\left[\frac{\mathcal{A}(R_1)}{4G}+\frac{\mathcal{A}(R_2)}{4G}-\frac{\mathcal{A}(R_1R_2)}{4G} \right] \nonumber \\
    &+ \left[S(\mathcal{E}(R_1))+S(\mathcal{E}(R_2))-S(\mathcal{E}(R_1R_2)) \right].
\end{align}
The $O(N^2)$ term in the mutual information undergoes a transition, from $0$ when the regions are small and far apart, to non-zero when regions are moved closer together or made larger. This is because for widely separated regions we have $\mathcal{A}(R_1R_2)=\mathcal{A}(R_1)\cup \mathcal{A}(R_2)$, and consequently the area terms in \ref{eq:mutualinfo} will cancel and give zero as illustrated in figure \ref{fig:disconnectedsurfaces}. For nearby or large regions the minimal surfaces take on the alternative configuration shown in \ref{fig:connectedsurfaces}, where $\mathcal{A}(R_1R_2)\neq \mathcal{A}(R_1)\cup \mathcal{A}(R_2)$, in which case there is a non-zero $O(N^2)$ term in the mutual information. 

The minimal surface perspective given by the RT formula and the central region perspective given by conjecture \ref{conj:entangledregions} both tie the transition in the mutual information from $O(1)$ to $O(N^2)$ to a geometric construction. We can combine these perspectives to arrive at statement relating the two geometric constructions, which we give below.
\begin{conjecture}\label{conj:geometricversion}
Consider two boundary regions $R_1$ and $R_2$ defined by $D(R_i) = \hat{J}^+(c_i)\cap \hat{J}^-(r_1)\cap \hat{J}^-(r_2)$, and suppose that the boundary central region is empty, $\hat{P}=\emptyset$. Then $\mathcal{A}(R_1R_2) = \mathcal{A}(R_1)\cup \mathcal{A}(R_2)$ implies that $P=J^-(r_1)\cap J^-(r_2) \cap J^+(c_1)\cap J^+(c_2) = \emptyset$.  
\end{conjecture}

To argue for this, suppose that $\mathcal{A}(R_1R_2) = \mathcal{A}(R_1)\cup \mathcal{A}(R_2)$. Then from the Ryu-Takayanagi formula, we have that
\begin{align}
    I(R_1:R_2) = \left[\frac{\mathcal{A}(R_1R_2)}{4G} - \frac{\mathcal{A}(R_1)}{4G} - \frac{\mathcal{A}(R_2)}{4G}\right] + O(N^0) = O(N^0). 
\end{align}
From theorem \ref{conj:entangledregions}, we have that $\hat{P}=\emptyset$ and $P\neq\emptyset \implies I(R_1:R_2) = O(N^2)$. Taking the contrapositive (and using that $\hat{P}=\emptyset$ by assumption), we get that $I(R_1:R_2)=O(N^0)\implies P=\emptyset$, as needed.

This statement is illustrated in figure \ref{fig:phasetransition3d}. It would be interesting to understand if this statement can be proven from a gravity perspective, and if assuming the above implies constraints on the stress tensor.

We can do an explicit check on conjecture \ref{conj:geometricversion} in the case of vacuum AdS$_{2+1}$. Consider the input and output points defined in equation \ref{eq:ads3example}. Then the region $D(R_1)$ is the spatial interval $(-x-\alpha,-\alpha)$ and $D(R_2)$ is the interval $(\alpha,\alpha+x)$, both on the $t=0$ time-slice. The regions $R_1$ and $R_2$ along with the input and output points are shown in figure \ref{fig:BB84task}. Considering the background geometry to be pure AdS, we can check when $P \neq \emptyset$. We can also explicitly check when the minimal surface enclosing $R_1R_2$ takes on the connected configuration of figure \ref{fig:connectedsurfaces}. In appendix \ref{app:minsurfaceandbulkpoint} we find that both $P\neq \emptyset$ and the minimal surface is connected exactly when
\begin{align}\label{eq:Pcondition}
    \sin^2(x/2) \geq \sin(x+\alpha)\sin(\alpha).
\end{align}
This confirms conjecture \ref{conj:geometricversion} for this example. While we made the assumption of regions of equal size, regions of unequal size may be brought to equal size by a conformal transformation, confirming conjecture \ref{conj:geometricversion} in those cases as well. We can also note that while conjecture \ref{conj:geometricversion} only gave the one way implication that $P = \emptyset \implies \mathcal{A}(R_1R_2) = \mathcal{A}(R_1)\cup \mathcal{A}(R_2)$, the implication turns out to run both ways in the case of vacuum AdS$_{2+1}$. We would not expect this to be the case generically, due to the one way implication in principle \ref{eq:AQTprinciple} along with the possibility of the mutual information counting a classical contribution to the correlation between $R_1$ and $R_2$.

\section{Holographic procedures}\label{sec:procedures}

In this section we give two methods for completing asymptotic quantum tasks in the boundary theory whenever they can be completed in the bulk theory: the AdS/CFT procedure inherited from the AdS/CFT dictionary, and the teleportation procedure developed in quantum cryptography. In both cases we study how efficiently the procedure replaces bulk geometry with entanglement. 

\subsection{The AdS/CFT procedure}\label{sec:adscftspoof}

Given a bulk procedure to complete a quantum task, the AdS/CFT dictionary provides a boundary description of the same task. The AdS/CFT dictionary is reviewed elsewhere \cite{aharony2000large,harlow2018tasi}, but we study some features of how the dictionary provides a procedure here.

As an example we consider a summoning task with the geometry shown in figure \ref{subfig:cylinder}. We consider the CFT to be in its vacuum state before the inputs are received. The input and output points are, in $(t,\varphi,\rho)$ coordinates,
\begin{align}
    c_1 &= (0,0,\infty) \,\,\,\,\,\,\,\,\,\,\,\,\,\,\,\,\,\,\,\,\,\,\,\, r_1 = (\pi,\pi/2,\infty) \nonumber \\
    c_2 &= (0,\pi,\infty) \,\,\,\,\,\,\,\,\,\,\,\,\,\,\,\,\,\,\,\,\,\,\, r_2 = (\pi,3\pi/2,\infty).
\end{align}
Notice that $\hat{P}=\hat{J}^+(c_1)\cap \hat{J}^+(c_2)\cap \hat{J}^-(r_1)\cap \hat{J}^-(r_2) = \emptyset$, while $P$ is not empty and consists of the point $(\pi/2,0,0)$. The inputs are $\ket{\psi}$ at $c_1$ and $b\in \{1,2 \}$ at $c_2$. To complete the task successfully Alice returns $\ket{\psi}$ at $r_b$. 

Considering first the limit of $N=\infty$, the bulk geometry becomes entirely classical. To complete the task in the bulk the inputs $\ket{\psi}$ and $b$ can be brought to $P$, and $\ket{\psi}$ then routed to the correct output point $r_b$. This completes the task with a perfect success rate. We can use the AdS/CFT dictionary to translate this bulk protocol into a boundary one. As $\ket{\psi}$ and $b$ fall deeper into the bulk, they are smeared over the boundary degrees of freedom and recorded into a holographic error correcting code. Entanglement wedge reconstruction informs us of which boundary regions are able to access $\ket{\psi}$ and $b$. In particular, it is possible to reconstruct the bulk degrees of freedom $\ket{\psi}$ or $b$ from a boundary region $R$ whenever $\ket{\psi}$ or $b$ lives in the entanglement wedge of $R$ \cite{dong2016reconstruction,cotler2017entanglement}. While at finite $N$ this reconstruction is approximate, at infinite $N$ it is exact. At time $t=\pi/2$, $\ket{\psi}$ and $b$ are in the center of AdS, and any half space of the boundary can be used to reconstruct $\ket{\psi}$. Indeed, looking at the projection of the backward light cones of $r_1$ and $r_2$ onto the $t=\pi/2$ slice we see that Alice will need to do just that: to complete the task she must reconstruct $\ket{\psi}$ from the interval $(0,\pi)$ if $b=1$ or from the interval $(\pi,2\pi)$ if $b=2$. Since we are at $N=\infty$ she may do so exactly. 

To phrase this in a way we can make robust, we imagine a quantum channel that completes the task perfectly, which we call $\mathcal{N}_{\text{ideal}}$. We label the actual implementation of the task as $\mathcal{N}$. At infinite $N$ we have
\begin{align}
    \mathcal{N}_{\text{ideal}} = \mathcal{N}.
\end{align}
At finite $N$ we expect this to be relaxed. To characterize this it is helpful to introduce the diamond norm, which can be used to construct a distance measure between quantum channels \cite{kitaev2002classical,wilde2013quantum}. The diamond-norm distance between two channels $\mathcal{N}^1_A$, $\mathcal{N}^2_A$ which act on a Hilbert space $A$ is defined by
\begin{align}
    ||\mathcal{N}^{1} - \mathcal{N}^2||_{\diamond} \equiv \max_{\Psi,R} |\mathcal{I}_R\otimes\mathcal{N}^{1}_A(\Psi_{RA}) - \mathcal{I}_R\otimes \mathcal{N}^2_A(\Psi_{RA})|_1,
\end{align}
where on the right we've employed the trace distance $| \rho-\sigma|_1 = \tr|\rho-\sigma|$. Notice that the maximization is over the choice of input density matrix $\Psi$ as well as over the choice of ancilla $R$, on which the channels act identically. The diamond-norm distance is operationally meaningful in that it determines the success probability in distinguishing two channels \cite{wilde2013quantum,pirandola2017ultimate}. 

We would like to understand how close Alice can come to applying the channel $\mathcal{N}_{\text{ideal}}$ when $N<\infty$. Note that the diamond distance between the ideal channel and any implemented channel is strictly positive for finite $N$. To see this, recall that the causal structure of the task requires that $\ket{\psi}$ be possible to reconstruct from both intervals $(0,\pi)$ and $(\pi,2\pi)$ of the $t=\pi/2$ slice, and that bulk reconstruction from boundary subregions becomes approximate at finite $N$. More precisely, the JLMS \cite{jafferis2016relative} result
\begin{align}
    \frac{C}{N} = S(\rho_{bulk}||\sigma_{bulk}) - S(\rho_{bnd}||\sigma_{bnd}),
\end{align}
with $C$ independent of $N$, straightforwardly gives that exact reconstruction is impossible at finite $N$. This is because the relative entropy is decreasing under quantum channels, so that for any channel $\mathcal{R}$,
\begin{align}
\frac{C}{N} &= S(\rho_{bulk}||\sigma_{bulk}) - S(\rho_{bnd}||\sigma_{bnd}) \nonumber \\
&\leq S(\rho_{bulk}||\sigma_{bulk}) - S(\mathcal{R}(\rho_{bnd})||\mathcal{R}(\sigma_{bnd})).
\end{align}
If there were a channel $\mathcal{R}$ which perfectly recovered bulk density matrices from boundary ones the above would be a contradiction. 

We can also argue that $||\mathcal{N}_{\text{ideal}}-\mathcal{N}||_{\diamond}$ should grow with $N$ no faster than $1/\sqrt{N}$. This follows under the optimistic assumption that recovering $\ket{\psi}$ approximately from a boundary interval is the largest contribution to the diamond distance\footnote{Another contribution could come from reconstructing $\ket{\psi}$ from the wrong interval, for instance.}. If we do so, we can employ the theory of universal recovery channels \cite{junge2018universal,cotler2017entanglement} to bound the diamond norm from above. A central result in the understanding of universal recovery channels is that if a channel $\mathcal{N}$ changes the relative entropy by only a small amount, then there exists a good inverse channel to $\mathcal{N}$. More precisely, if we let $\mathcal{N}$ be the bulk to boundary map so that $\mathcal{N}(\rho_{\text{bulk}}) = \rho_{\text{bnd}}$, we have that there exists a channel $\mathcal{R}_{\sigma,\mathcal{N}}$ such that
\begin{align}
S(\rho_{\text{bulk}}||\sigma_{\text{bulk}}) - S(\rho_{\text{bnd}}||\sigma_{\text{bnd}}) \geq -\log F(\rho_{\text{bulk}},\mathcal{R}_{\sigma,\mathcal{N}}(\rho_{\text{bnd}})).
\end{align}
Using that the left hand side is of order $1/N$ and rearranging we get that
\begin{align}
    e^{-C/N} \leq F(\rho_{\text{bulk}},\mathcal{R}_{\sigma,\mathcal{N}}(\rho_{\text{bnd}})).
\end{align}
From the bulk density matrix Alice may extract the qubit holding $\ket{\psi}$ by applying an appropriate quantum channel. Since the fidelity increases under quantum channels
\begin{align}\label{eq:fidelityboundfromN}
    e^{-C/N} \leq F(\psi,\psi_R),
\end{align}
where $\psi$ is the density matrix $\ketbra{\psi}{\psi}$ and $\psi_R$ is the recovered approximation to $\psi$.

Finally, we translate our bound on the fidelity to a bound on the distance between the ideal and implemented protocols. We can bound the diamond norm by employing the standard inequality \cite{fuchs1999cryptographic}
\begin{align}
    |\rho-\sigma|_1 \leq \sqrt{1-(F(\rho,\sigma))^2},
\end{align}
along with inequality \ref{eq:fidelityboundfromN},
\begin{align}
    ||\mathcal{N}_{\text{ideal}} - \mathcal{N}||_{\diamond} &\leq \max_{\Psi,R} \sqrt{1 - (F(\mathcal{I}_R\otimes \mathcal{N}_{ideal}(\Psi),\mathcal{I}_R\otimes \mathcal{N}(\Psi)))^2} \nonumber \\
    &\leq \sqrt{1 - (F(\psi,\psi_R))^2} \nonumber \\
    &\leq \sqrt{1- e^{-C/N}} \nonumber \\
    &\leq \sqrt{\frac{C}{N}},
\end{align}
where we absorbed a factor of two into the definition of the constant $C$. We can summarize then by saying
\begin{align}\label{eq:diamondnorm1}
    0\leq ||\mathcal{N}_{\text{ideal}} - \mathcal{N}||_{\diamond} \leq \frac{C}{\sqrt{N}},
\end{align}
with the equality on the left holding if and only if $N=\infty$, and we've derived the upper bound only under the assumption that reconstructing the state $\ket{\psi}$ from an interval is the main source of error in the protocol. Although we've studied one particular quantum task, we expect that these bounds on $||\mathcal{N}_{\text{ideal}} - \mathcal{N}||_{\diamond}$ are generic whenever we consider finite $N$. 

We have arrived at \ref{eq:diamondnorm1} by considering the boundary protocol. However, according to our principle \ref{thm:AQT} if $\mathcal{N}$ can be applied in the boundary it can be applied in the bulk, so the same closeness result must apply to a bulk implementation of the protocol. From the bulk perspective it is less clear how to understand the source of the approximation, but we can plausibly attribute it to stringy corrections at finite $N$. 

We can relate the parameter $N$ to the mutual information $I$ between the two boundary intervals relevant to the task. We regulate the mutual information by introducing a small separation between the two intervals. The regularized mutual information will then be finite and scale with $N$ according to $I\sim N^2$. After sending the regulator to zero the two intervals form a partition of the boundary, so this is the mutual information of two parts of a pure state and represents a measure of entanglement between the two regions. Equation \ref{eq:diamondnorm1} becomes
\begin{align}\label{eq:sum1}
    0 \leq ||\mathcal{N}_{\text{ideal}} - \mathcal{N}||_{\diamond} \leq \frac{C}{I^{1/4}}.
\end{align}
To interpret this result recall that $\mathcal{N}_{\text{ideal}}$ can be completed perfectly at $N=\infty$, where the bulk becomes entirely classical. $\mathcal{N}$ is an approximation to this when $N$ becomes finite, which coincides with the amount of entanglement in the problem becoming finite (after the UV part has been regulated). The distance between channels above then is a probe of how well classical geometry may be simulated with a finite amount of entanglement. 

\subsection{The teleportation procedure}\label{sec:teleportationspoof}

There are two general strategies in the quantum cryptography literature \cite{buhrman2014position,beigi2011simplified} for completing asymptotic quantum tasks in the boundary. We will focus on the port-teleportation based procedure given in \cite{beigi2011simplified} since it is conceptually simpler. For simplicity we describe the procedure in the case of two input and two output points. 

To begin constructing the port-based procedure it is helpful to consider a naive and incorrect strategy. Suppose that systems $A_1$ and $A_2$ are input at $\mathrm{c}_1$ and $\mathrm{c}_2$ respectively, and Alice's goal is to apply a channel $\mathcal{N}:A_1A_2\rightarrow B_1B_2$ before outputting the $B_1$ and $B_2$ systems at $r_1$ and $r_2$. The naive strategy is to teleport $A_1$ onto a system $A_1'$ held near $\mathrm{c}_2$, apply the channel $\mathcal{N}$, then send each $B_i$ to the corresponding $r_i$. Unfortunately, this naive strategy will usually fail. If the state on $A_1A_2$ was $\ket{\psi}$ before the teleportation, then afterwards the state on $A_1'A_2$ is
\begin{align}
    \mathcal{P}^i_{A_1'} \otimes \mathcal{I}_{A_2} \ket{\psi},
\end{align}
where $\mathcal{P}^i$ is a randomly chosen Pauli operator acting on each of the qubits in $A_1$. If $A_1$ consists of $n$ qubits, then with probability $1-1/4^n$ the state on the system $A_1'A_2$ is changed by a non-trivial operator acting on $A_1'$. Since in general the Pauli operators will not commute with the operation $\mathcal{N}$ the procedure fails with high probability. 

\begin{figure}
    \centering
    \begin{tikzpicture}
    
    \draw[fill=lightgray] (0,0) -- (0,2) -- (2,2) -- (2,0) -- (0,0);
    \node at (1,1) {\Large{$\Pi_i$}};
    
    \draw (0.25,-0.5) -- (0.25,0);
    \node[below] at (0.25,-0.5) {$\ket{\psi}_A$};
    
    \draw[blue,->] (1,2) -- (1,2.5);
    \node[above right] at (0.8,2.5) {$i_*\in \{1,...,M\}$};
    
    \draw (1.75,0) -- (3.75,-1) -- (5.75,0) -- (5.75,0.25);
    \draw (1.5,0) -- (3.75,-1.15) -- (6,0) -- (6,0.25);
    \draw (1.25,0) -- (3.75,-1.3) -- (6.25,0) -- (6.25,0.25);
    \draw (1,0) -- (3.75,-1.45) -- (6.5,0) -- (6.5,0.25);
    
    \draw (4.5,-0.5) -- (5,-1);
    \node[below] at (5,-1) {$M$};
    
    \node[below] at (3.75,-1.45) {$\ket{\Psi^+}_{B_1^iB_2^i}^{\otimes M}$};
    
    \draw[gray,->] (7,0.75) to [out=180,in=90] (6.25,0.25);
    \node at (7,0.75) [gray,align=left,right]{\small{With high fidelity,} $\ket{\psi}$ \\ \small{appears on the $B_{i_*}$ system.}};
    
    
    \end{tikzpicture}
    \caption{The port-teleportation protocol. A state $\ket{\psi}$ is held in system $A$, along with $M$ entangled systems $\ket{\Psi_A}_{B_1^iB_2^i}^{\otimes M}$ where each $\ket{\Psi_A}$ consists of $n$ EPR pairs, where $n$ is the number of qubits in $A$. The $B_2^i$ are referred to as ``ports''. A measurement with measurement operators $\{\Pi_i\}$ is performed on the $A B_1^1B_1^2...B_1^M$ system producing output $i_*\in \{1,2...,M\}$. The state $\ket{\psi}$ then appears on the $B_2^{i_*}$ system with a fidelity controlled by $1/M$.}
    \label{fig:port_teleport}
\end{figure}
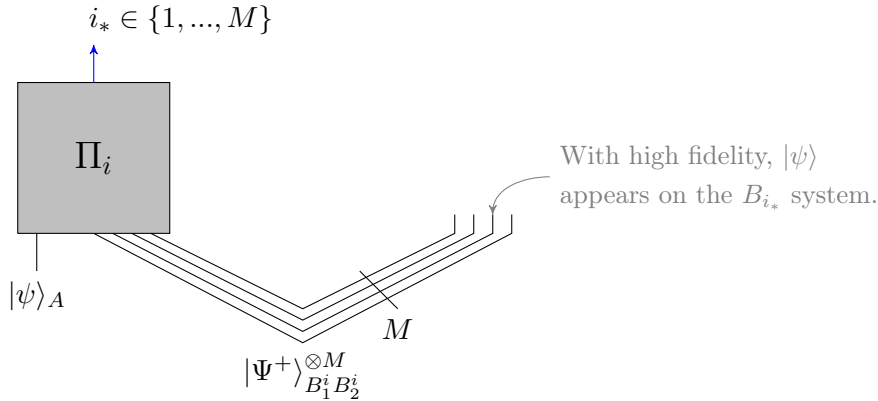

Adding a port-teleportation \cite{ishizaka2008asymptotic} to this protocol allows us to get around this difficulty. We illustrate the functionality of port-teleportation in figure \ref{fig:port_teleport}. Port-teleportation shares the basic features of standard teleportation. The protocol involves a system $A$ on which the state $\ket{\psi}$ to be teleported is stored, and an entangled system $X_1X_2$ used as a resource for the teleportation. A joint measurement is performed in the $AX_1$ system, then the measurement outcome is broadcast and combined with $X_2$ to reproduce $\ket{\psi}$ again. In any teleportation scheme compatibility with causality requires the $X_2$ target system to reveal no information about the teleported state until the classical data has reached it. In traditional teleportation this is achieved by the appearance of random Pauli operators on the $X_2$ system. Port-teleportation satisfies this causality constraint differently. The $X_2$ system is much larger than the system holding the state to be teleported. In fact $X_2 = X_2^1,...,X_2^M$ with $\dim A = \dim X_2^i$ for each $i$ and $M \gg 1$. Once the measurement is performed in the port-teleportation protocol the state, untampered except for a small perturbation, appears on one of the output ports $X_2^1,...,X_2^M$. The classical measurement outcome obtained in performing the port-teleportation reveals which port the state appears on. The size of the small perturbation and the number of ports are such that no information is revealed about the teleported state before the classical data arrives; the size of the small perturbation may be diminished as much as desired, at the expense of adding additional ports. In particular if $M$ ports are used the state appears on one of the ports with a fidelity 
\begin{align}\label{eq:portfidelity}
    f = 1 - \frac{d^2-1}{4M}
\end{align}
where $d$ is the dimension of the teleported system. 

Unlike traditional teleportation in which the teleported state may, in principle, be reproduced exactly on the target qubit, port-teleportation is necessarily approximate, at least in the finite dimensional setting. One way to understand why this must be the case is by noting that a port-teleportation scheme with perfect fidelity could be used to construct a universal quantum processor using only a finite dimensional system \cite{ishizaka2008asymptotic}, a feat known to be impossible \cite{nielsen1997programmable}.

We can use port teleportation to build on our earlier naive protocol. After the first teleportation of $A_1$ from Alice$_1$ to Alice$_2$, Alice$_2$ port teleports $A_1'A_2$ back to Alice$_1$. Alice$_1$ knows which Pauli operator appears on $A_1'$, but not on which port the state has appeared. This is not a problem however, as she may simply apply the correcting Pauli operator to every port. She then knows she holds the joint state $\ket{\psi}$ on one of her ports, but not which one, so she again applies the needed channel to every port. She takes all of the outputs from each application of the channel and sends them to the appropriate output points. Meanwhile, Alice$_2$ has obtained a measurement outcome which reveals which port the state actually appeared on. She sends this measurement outcome to both output points. Near the output points Alice discards all but the outputs from the correct port. 

To describe this protocol in more detail it is useful to define the following states. We define the maximally entangled state
\begin{align}
    \ket{\Psi^+} = \frac{1}{\sqrt{2}}\left(\ket{00}+\ket{11} \right).
\end{align}
The system input to Alice in the task will be $\ket{\psi}_{A_1A_2C}$. $A_1$ is received at $c_1$, $A_2$ at $c_2$, and $C$ is a purifying system held by Bob. We will assume $\dim A_1=\dim A_2$ (if not, add ancillas to the smaller system). We will define a state $\ket{\Psi_{A_1}}$ as 
\begin{align}
    \ket{\Psi_{A_1}} = \ket{\Psi^+}^{\otimes n}
\end{align}
where $n$ is the number of qubits in $A_1$, so that $\ket{\Psi_A}$ is large enough to teleport the $A_1$ system. Finally, we define the state
\begin{align}
    \ket{\Psi_{A_1A_2,M}} = \ket{\Psi_{A_1A_2}}^{\otimes M}.
\end{align}
This is an entangled state large enough to perform the port teleportation protocol with $M$ ports on the $A_1A_2$ system.

With these definitions we are now ready to give a formal description of the protocol:
\begin{enumerate}
    \item At an early time Alice distributes the entangled states $\ket{\Psi_{A_1}}_{YA_1'}$ and $\ket{\Psi_{A_1A_2,M}}_{X_1X_2}$ between the spatial locations of $c_1$ and $c_2$. $X_1$ and $X_2$ are divided into $M$ subsystems of the same size as $A_1A_2$.
    \item After receiving the $A_1$ system Alice$_1$ teleports, using the standard teleportation procedure, the $A_1$ system onto the $A_1'$ system using the state $\ket{\Psi_{A_1}}_{YA_1'}$. She obtains a measurement outcome $i_*$. The $A_1'A_2C$ system is now in the state
    \begin{align}
        \mathcal{P}^{i_*}_{A_1'}\otimes \mathcal{I}_{A_2C}\ket{\psi}_{A_1'A_2C}.
    \end{align}
    and is held by Alice$_2$.
    \item Alice$_2$ teleports $A_1'A_2$ to Alice$_1$ using the port based protocol and the state $\ket{\Psi_{A_1A_2,M}}_{X_1X_2}$. She obtains a measurement outcome $j_*$. Alice$_1$ holds systems $X_1=X_1^1X_1^2...X_1^M$, and the state on $X_1^{j_*}$ is
    \begin{align}
        P^i_{X_1^{j_*}} \ket{\psi'} \approx P^i_{X_1^{j_*}} \ket{\psi}_{X_1^{j_*}C}
    \end{align}
    where $P^{i}= \mathcal{P}^{i_*}\otimes \mathcal{I}$, and the closeness of the approximation is controlled by $1/M$.
    \item Alice$_1$ applies $(P^i)^{-1}$ to each of the $X_1^i$ systems, so that the $X_1^{j_*}C$ system is
    \begin{align}
        \ket{\psi'}_{X_1^{j_*}C} \approx \ket{\psi}_{X_1^{j_*}C}
    \end{align}
    \item Alice$_1$ applies the quantum channel $\mathcal{N}$ to each of the subsystems $X_1^j$. The $X_1^{j_*}C$ system is now in the state
    \begin{align}
        \mathcal{N}_{X_1^{j_*}}\otimes \mathcal{I}_C(\ketbra{\psi'}{\psi'})_{X_1^{j_*}C} \approx \mathcal{N}_{X_1^{j_*}}\otimes \mathcal{I}_C(\ketbra{\psi}{\psi})_{X_1^{j_*}C}
    \end{align}
    The channel $\mathcal{N}_{X_1^{j_*}}$ maps the $X_1^{j_*}$ system to the two output systems $B_1^{j_*}$ and $B_2^{j_*}$. 
    \item Alice$_1$ sends the $B_1^j$ systems to $\mathrm{r}_1$ and the $B_2^j$ systems to $\mathrm{r}_2$. 
    \item Alice$_2$ sends her measurement outcome $j_*$ to both $\mathrm{r}_1$ and $\mathrm{r}_2$.
    \item At $\mathrm{r}_1$,  Alice$_1$ discards all the $B_1^j$ systems except $B_1^{j_*}$, which she returns to Bob. Similarly, Alice$_2$ discards all the $B_2^j$ except $B_2^{j_*}$, which she returns to Bob. 
\end{enumerate}
We refer the reader to \cite{beigi2011simplified} for further details on this protocol. 

The number of EPR pairs used in the port-teleportation holographic procedure depends on the number of qubits $n$ required to hold the $A_1$ system (we assume $\dim A_1=\dim A_2$), and on the parameter $M$, which controls how closely we simulate applying the channel $\mathcal{N}$ to the $A_1A_2$ systems. Both of these determine the total number of EPR pairs $E$ used in the protocol. Call the channel we apply via the holographic procedure $\mathcal{N}$ and the asked for channel $\mathcal{N}_{\text{ideal}}$. Then a careful analysis \cite{beigi2011simplified} of the protocol given above reveals that the distance between the ideal and implemented channels scales with $n$ and the number of EPR pairs $E$ used in the protocol according to
\begin{align}\label{eq:diamondbound2}
    ||\mathcal{N}_{\text{ideal}} - \mathcal{N}||_{\diamond} \leq \frac{\sqrt{n}2^{4n + 5/2}}{\sqrt{E}}.
\end{align}
We can interpret this statement as follows. The channel $\mathcal{N}_{\text{ideal}}$ is one we'd be able to do if given access to a classical bulk geometry. $\mathcal{N}$ is a channel we can do using $E$ EPR pairs and never accessing the bulk region. The bound \ref{eq:diamondbound2} expresses how well a given number of EPR pairs can be used to replace the classical bulk region. 

To compare this to the analogous result \ref{eq:sum1} for AdS/CFT, we should rewrite equation \ref{eq:diamondbound2} in terms of a mutual information. The relevant state to consider is the entangled state which is shared between Alice$_1$ and Alice$_2$ at the start of the protocol. This mutual information scales linearly with the number of EPR pairs $E$. Then \ref{eq:diamondbound2} becomes
\begin{align}\label{eq:sum2}
    ||\mathcal{N}_{\text{ideal}} - \mathcal{N}||_{\diamond} \leq \frac{C'}{I^{1/2}},
\end{align}
where $C'$ is an $I$ independent number. We see that the teleportation procedure approximates classical geometry more efficiently than the AdS/CFT procedure, at least when fixing the number of input qubits $n$. 

\section{Discussion}\label{sec:discussion}

Our starting point has been that \emph{an information processing task can be accomplished in the bulk exactly when it can be accomplished in the boundary}. Importantly, we have been careful when drawing conclusions about the low energy effective bulk description to only use this implication from the bulk to the boundary, since the reverse implication generally takes us outside the bulk low energy effective theory. We have further refined this principle by considering information processing tasks with inputs and outputs distributed throughout spacetime. To be able to canonically identify tasks in the bulk with tasks in the boundary, we focused on asymptotic quantum tasks, where the inputs are located on the spacetime boundary. Our guiding principle can be stated then as
\begin{align}
    \text{AQT possible as bulk task} \implies \text{AQT possible as boundary task}.
\end{align}
By starting with a choice of geometry and considering AQTs that occur in that spacetime, we could deduce certain features from this principle of the corresponding holographic theory. 

The AQT perspective suggests a surprising connection between entanglement and bulk causal structure. In particular, we argued that a region $P$ being non-empty implies a mutual information is order $N^2$, where $P$ is formed from the intersection of four light cones. This connection could be deduced using very little about the boundary theory --- we used only that the boundary theory is quantum mechanical and obeys relativistic causality. This result gives an operational perspective on why the entanglement structure of holographic theories is controlled by the entanglement wedge, and why the mutual information should undergo a phase transition. When combined with the Ryu-Takayanagi formula, we obtain a relation between properties of minimal surfaces and bulk causal structure. 

In the context of this connection between minimal surfaces and causal structure, there are various directions that could be pursued. Recall that we proved that a non-empty empty central bulk region implies a positive mutual information, but not the reverse implication. It is natural to ask if the implication may run in both directions (as it happened to in the vacuum case), and if so whether or not the geometry of the bulk central region might determine the value of the mutual information quantitatively. In the future we intend to study geometries with matter present and check if the central region being non-empty and mutual information being positive coincide exactly in explicit examples. It would also be interesting to study the purely geometric statement given as conjecture \ref{conj:geometricversion} from a gravity viewpoint. Plausibly, the truth of this theorem implies constraints on the stress tensor. Since this gravitational theorem was proven using the existence of a boundary dual theory, these would be interpreted as constraints that must be satisfied for the gravitational theory to have a consistent boundary description. This is similar to the constraints derived from strong subadditivity or other entropic inequalities \cite{banerjee2014constraining,lin2014tomography,banerjee2015nonlinear,lashkari2015inviolable,bhattacharya2015entanglement,lashkari2016gravitational,neuenfeld2018positive}. 

The AQT perspective also raised the notion of holographic procedures, which are methods of replacing bulk classical geometry with boundary entanglement. Results in quantum cryptography provide one holographic procedure, while AdS/CFT provides another. In both procedures asymptotic quantum tasks that can be achieved perfectly using access to a classical bulk region can be completed only approximately, with the closeness of the approximation controlled by the amount of available entanglement. Specifically the distance between the intended and achieved channels, measured using the diamond norm, scales like $I^{-1/2}$ in the cryptographic case and $I^{-1/4}$ in the AdS/CFT case, with $I$ the mutual information between two regions relevant to the problem. It would be interesting to study holographic procedures in a more general setting and understand what, if any, are the fundamental limits on how efficiently classical geometry may be simulated using entanglement. 

In another direction, it would be interesting to understand if the AdS/CFT holographic procedure could be applied usefully to cryptography. One could try and apply the AdS/CFT holographic procedure as a spoofing scheme. Unfortunately the AdS/CFT procedure does not carry immediately over to the cryptographic setting, where tasks occur in a Minkowski space background. Nonetheless some of the general ideas of AdS/CFT may be useful in the cryptographic context, for instance holographic error correcting codes \cite{pastawski2015holographic}.

Using the language of relativistic quantum tasks allows connections to be drawn between holography and a body of literature within quantum information theory concerned with relativistic quantum tasks. The study of relativistic quantum tasks is in its infancy however, and even some relatively simple tasks have evaded a full characterization\footnote{For instance, the ``spooky summoning'' tasks \cite{adlam2017thesis}.}. The connection between quantum tasks and holography discussed here adds a new motivation for the study of quantum tasks. While we have drawn on the existing quantum tasks literature, it would be interesting to understand what further results on tasks would be of the deepest interest to the holographer.

\section{Acknowledgements}

I thank David Wakeham, Dominik Neuenfeld, Mark Van Raamsdonk, Patrick Hayden, Geoffrey Pennington, Jonathan Sorce, and Jason Pollack for useful discussions. I acknowledge support from the It from Qubit Collaboration, which is sponsored by the Simons Foundation. I was also supported by a CGS-D award given by the National Research Council of Canada.

\appendix

\section{Proof of necessity of entanglement for the \texorpdfstring{$\mathbf{\hat{B}}_{84}^{\times n}$}{TEXT}  task}\label{app:necessityproof}

In this appendix we prove theorem \ref{thm:necessityofentanglement}, which bounds the success probability for the $\mathbf{B}_{84}^{\times n}$ task when the in-regions share zero mutual information. Appendix \ref{app:linearbound} relaxes this requirement, and bounds the parameter $n$ (the number of times the task may be repeated in parallel) in terms of the mutual information. Our proof of theorem \ref{thm:necessityofentanglement} follows \cite{tomamichel2013monogamy}, which developed results on a monogamy of entanglement game and subsequently applied them to $\mathbf{B}_{84}^{\times n}$. We introduce the monogamy of entanglement game and apply it's analysis to $\mathbf{B}_{84}^{\times n}$. For the technical analysis of the monogamy game itself we refer the reader to the original article.

We repeat theorem \ref{thm:necessityofentanglement} here for convenience. \\

\noindent \textbf{Theorem} \ref{thm:necessityofentanglement}
\emph{Consider the $\mathbf{\hat{B}}_{84}^{\times n}$ task, with input points $c_1,c_2$ and output points $r_1,r_2$. Then if the central region $\hat{P}$ is empty and $I(R_1:R_2)=0$, then the task can be completed with probability at most $\beta^n$ where $\beta = \frac{1}{2}+ \frac{1}{2\sqrt{2}}\approx 0.85$.}
\vspace{0.4cm}

The monogamy game\footnote{In fact \cite{tomamichel2013monogamy} discuss a more general class of monogamy games than described here. For simplicity we have taken the relevant special case.} is played among three players, Alice, Bob and Charlie and consists of three phases.
\begin{itemize}
    \item \textbf{Preparation phase}: Bob and Charlie prepare an arbitrary quantum state $\rho_{ABC}$. They then send system $A$ to Alice, where $A$ consists of $n$ qubits. Bob holds $B$ and Charlie holds $C$. Once this is done, Bob and Charlie may no longer communicate.
    \item \textbf{Question phase}: Alice chooses a random binary string of length $n$, call it $\theta=\theta_1...\theta_n$. She then measures her $i$th qubit in basis $\theta_i$, where $\theta=0$ corresponds to the computational basis and $\theta_i=1$ corresponds to the Hadamard basis. Alice then announces $\theta$ to Bob and Charlie.
    \item \textbf{Answer phase}: Bob and Charlie each act on $B$ and $C$ respectively to form independent guesses of Alice's measurement outcomes. 
\end{itemize}
If both Bob and Charlie guess all of the outcomes correctly, then we say Bob and Charlie won the game. 

The reason for calling this a monogamy game may already be clear: one of Bob or Charlie may always guess correctly by sharing maximally entangled states with Alice. For instance by sharing $\rho_{ABC} = \ketbra{\Psi^+}{\Psi^+}^{\otimes n}_{AB} \otimes \rho_C$, Bob may always guess correctly. To do this, after Alice announces her measurement bases, Bob simply measures $B$ in the same bases. Because entanglement is monogamous however, only one of Bob or Charlie may execute this strategy, and their success probability is bounded somewhere below one. 

The main result on this monogamy of entanglement game that we will make use of is the following.
\begin{theorem}\label{thm:monogamygame}
The monogamy of entanglement game played with $n$ qubits has a maximal success probability $p_{suc} = \beta^n$
where $\beta=\frac{1}{2}+\frac{1}{2\sqrt{2}}\approx 0.85$.
\end{theorem} 
We refer the reader to \cite{tomamichel2013monogamy} for the proof.

In order to apply this to the $\mathbf{\hat{B}}_{84}^{\times n}
$ task, it is convenient to rephrase the task in an alternative way. This is usually called the purified protocol, and runs as follows. 
\begin{itemize}
    \item At an early time Bob prepares maximally entangled states $\ket{\Psi^+}_{C_iA_{1i}}$.
    \item At $c_1$ Bob hands to Alice the $A_{1i}$ systems, and at $c_2$ Bob hands to Alice classical bits $q_i$.
    \item Bob measures the $C_i$ system in the $q_i$ basis, where $q_i=0$ corresponds to the computational basis and $q_i=1$ to the Hadamard basis. He obtains outcome $b_i$.
    \item Alice hands to Bob her guesses $b_{1i}$ at $r_1$ and $b_{2i}$ at $r_2$.
    \item If Bob's measurement outcome $b_{i}$ matches both $b_{1i}$ and $b_{2i}$ for all $i$, we say Alice has completed the task successfully. 
\end{itemize}
After Bob's measurement the $A_i$ system is in the state $H^{q_i}\ket{b_i}$. Thus, if Bob had measured before handing Alice the $A_1$ system, this would be exactly the original $\mathbf{\hat{B}}_{84}^{\times n}$ task. Alice's success probability cannot depend on when Bob makes this measurement however, since the system $C_i$ he measures is isolated from all of Alice's inputs. Consequently a bound on the success probability of the purified protocol will also apply to the original unpurified protocol. Since completing the purified task is equivalent to completing the monogamy of entanglement game, the bound in theorem \ref{thm:monogamygame} applies and theorem \ref{thm:necessityofentanglement} is proven. 

\section{Linear bound on the mutual information in the \texorpdfstring{$\mathbf{\hat{B}}_{84}^{\times n}$}{TEXT} task}\label{app:linearbound}

In this appendix we prove theorem \ref{thm:linearbound}. The proof strategy is to observe that the protocol used to complete $\mathbf{\hat{B}}_{84}^{\times n}$ takes as input both the inputs from Bob, $\mathcal{A}$, and additionally whatever resource state $\rho_{R_1R_2}$ Alice has available to her. We will be able to relate Alice's success probability in the task to her ability to distinguish a state $\rho_{R_1R_2}$ which completes the task with high probability from the product state $\rho_{R_1}\otimes \rho_{R_2}$, which by the theorem \ref{thm:necessityofentanglement} completes the task with low probability.

The proof uses a a few inequalities and distances on density operators which we summarize briefly here. The trace distance is defined by
\begin{align}
    ||\rho-\sigma||_1 = \tr|\rho-\sigma| 
\end{align}
the trace distance has an important operational interpretation in terms of the probability of successfully distinguishing two density matrices $\rho$ and $\sigma$, 
\begin{align}\label{eq:distuishboundtrace}
    \frac{1}{2} + \frac{1}{4}||\rho-\sigma|| \geq p_{dist}
\end{align}
Another important quantity for us is the fidelity, defined by
\begin{align}
    F(\rho,\sigma) = \left[\tr\sqrt{\sqrt{\rho}\sigma\sqrt{\rho}}\right]^2.
\end{align}
The trace distance and fidelity are related by \cite{fuchs1999cryptographic,wilde2013quantum}
\begin{align}\label{eq:traceandfidelity}
    \frac{1}{2}||\rho-\sigma||_1 \leq \sqrt{1-F(\rho,\sigma)}.
\end{align}
Finally, the relative entropy and fidelity are related by
\begin{align}
    S(\rho||\sigma) \geq -2\log F(\rho,\sigma).
\end{align}
With these technical preliminaries in hand we are ready to give our proof of theorem \ref{thm:linearbound}.\\

\noindent \textbf{Theorem} \ref{thm:linearbound}:\emph{
For any state $\rho_{R_1R_2}$ which suffices to complete the $\mathbf{B}_{84}^{\times n}$ task with success probability $1$, the mutual information $I(R_1:R_2)$ is bounded below according to
\begin{align}\label{eq:linearbound}
\frac{1}{2}I(R_1:R_2) \geq  n \left(\log_2 1/\beta \right) - 1
\end{align}
where $\beta = \frac{1}{2}+ \frac{1}{2\sqrt{2}}\approx 0.85$ so that $\log_21/\beta \approx 0.23$.}\\

\begin{proof}
Recall that the mutual information can be expressed in terms of the relative entropy,
\begin{align}\label{eq:mutualinforrelativenetropy}
    I(R_1:R_2)_\rho = S(\rho_{R_1R_2}||\rho_{R_1}\otimes \rho_{R_2}).
\end{align}
The relative entropy is a measure of how distinguishable its two arguments are. To show we have a large mutual information then, we need to show $\rho_{R_1R_2}$ and $\rho_{R_1}\otimes \rho_{R_2}$ are easily distinguished.

Suppose that Alice is handed either the state $\rho_{R_1R_2}$ or  state $\rho_{R_1}\otimes \rho_{R_2}$. By assumption $\rho_{R_1R_2}$ allows the $\textbf{B}_{84}^{\times n}$ task to be completed with success probability $1$. From theorem \ref{thm:necessityofentanglement}, the success probability for any product state is bounded above by $\beta^n$. That is
\begin{align}\label{eq:pbounds}
    p_{suc}(\mathbf{\hat{B}}_{84}^{\times n},\rho_{R_1R_2})&=1, \nonumber \\
    p_{suc}(\mathbf{\hat{B}}_{84}^{\times n},\rho_{R_1}\otimes \rho_{R_2}) &\leq \beta^n.
\end{align}
Given this gap in success probabilities, Alice can use the $\textbf{B}_{84}$ task as a means to distinguish between $\rho_{R_1R_2}$ and $\rho_{R_1}\otimes \rho_{R_2}$. We imagine a scenario where Alice receives $\rho_{R_1R_2}$ with probability $1/2$ and $\rho_{R_1}\otimes \rho_{R_2}$ with probability $1/2$. She uses whichever state she receives as a resource state in completing the $\textbf{B}_{84}^{\times n}$ task. Alice then declares the state to be $\rho_{R_1R_2}$ if the task succeeds, and to be $\rho_{R_1}\otimes \rho_{R_2}$ otherwise. Alice's probability of correctly identifying the state will then be
\begin{align}
    p_{dist} = \text{Prob}&(\rho_{R_1R_2} \,\text{is received})\times p_{suc}(\mathbf{\hat{B}}_{84}^{\times n},\rho_{R_1R_2}) \nonumber \\
    +& \text{Prob}(\rho_{R_1}\otimes \rho_{R_2} \,\text{is received}) \times (1-p_{suc}(\mathbf{\hat{B}}_{84}^{\times n},\rho_{R_1}\otimes \rho_{R_2})).
\end{align}
Using \ref{eq:pbounds} and that each state is received with probability $1/2$ this becomes
\begin{align}
    p_{dist} \geq \frac{1}{2} + \frac{1}{2}(1-\beta^n)
\end{align}
The upper bound \ref{eq:distuishboundtrace} on the probability of distinguishing two states gives that
\begin{align}
    \frac{1}{2} + \frac{1}{4}||\rho_{R_1R_2}-\rho_{R_1}\otimes \rho_{R_2}||_1 \geq p_{dist}
\end{align}
so that we can relate the trace distance to $\beta^n$, 
\begin{align}
    \frac{1}{2}||\rho_{R_1R_2}-\rho_{R_1}\otimes \rho_{R_2}||_1 \geq 1-\beta^n.
\end{align}
Next, we employ the upper bound \ref{eq:traceandfidelity} on the trace distance in terms of the fidelity,
\begin{align}
    \sqrt{1-F(\rho_{R_1R_2},\rho_{R_1}\otimes \rho_{R_2})} \geq 1-\beta^n
\end{align}
which leads to
\begin{align}\label{eq:fbound}
    F(\rho_{R_1R_2},\rho_{R_1}\otimes \rho_{R_2}) \leq 2\beta^n.
\end{align}
Finally, we insert this bound on fidelity into the lower bound on relative entropy, and use equation \ref{eq:mutualinforrelativenetropy} to find
\begin{align}
    I(R_1:R_2)_\rho \geq -2\log F(\rho_{R_1R_2},\rho_{R_1}\otimes \rho_{R_2}) \geq -2\log (2\beta^n)
\end{align}
which proves the theorem. 
\end{proof}

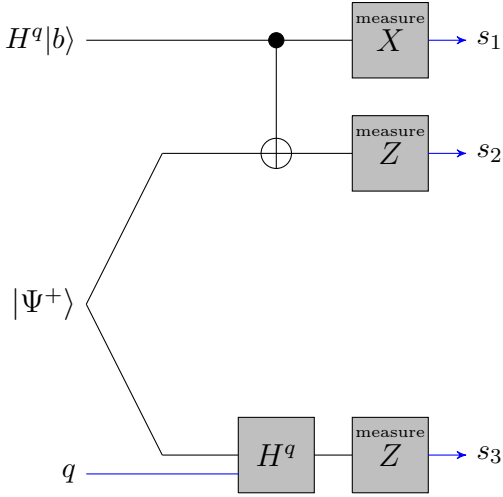
\begin{figure}
    \centering
    \begin{tikzpicture}[scale=1]
    
    \draw (0,0) -- (1,2);
    \draw (0,0) -- (1,-2);
    \node[left] at (0,0) {\large{$\ket{\Psi^+}$}};
    
    \draw (1,-2) -- (2,-2);
    
    \draw[fill=lightgray] (2,-2.5) -- (2,-1.5) -- (3,-1.5) -- (3,-2.5) -- (2,-2.5); 
    \node at (2.5,-2) {\large{$H^q$}};
    
    \draw (3,-2) -- (3.5,-2);
    
    \draw[fill=lightgray] (3.5,-2.5) -- (3.5,-1.5) -- (4.5,-1.5) -- (4.5,-2.5) -- (3.5,-2.5);
    \node[above] at (4,-1.9) {\tiny{measure}};
    \node at (4,-2) {\large{$Z$}};
    
    \draw (1,2) -- (3.5,2);
    \draw (0,3.5) -- (3.5,3.5);
    
    \node[left] at (0,3.5) {$H^q\ket{b}$};
    
    \draw[fill=black] (2.5,3.5) circle (0.1);
    \draw (2.5,3.5) -- (2.5,1.8);
    \draw (2.5,2) circle (0.2);
    
    \draw[fill=lightgray] (3.5,4) -- (4.5,4) -- (4.5,3) -- (3.5,3) -- (3.5,4);
    \node[above] at (4,3.6) {\tiny{measure}};
    \node at (4,3.5) {\large{$X$}};
    
    \draw[fill=lightgray] (3.5,2.5) -- (4.5,2.5) -- (4.5,1.5) -- (3.5,1.5) -- (3.5,2.5);
    \node[above] at (4,2.1) {\tiny{measure}};
    \node at (4,2) {\large{$Z$}};
    
    \draw[blue,->] (4.5,2) -- (5,2);
    \node[right] at (5,2) {$s_2$};
    
    \draw[blue,->] (4.5,3.5) -- (5,3.5);
    \node[right] at (5,3.5) {$s_1$};
    
    \draw[blue,->] (4.5,-2) -- (5,-2);
    \node[right] at (5,-2) {$s_3$};
    
    \draw[blue] (0,-2.25) -- (2,-2.25);
    \node[left] at (0,-2.25) {$q$};
    
    \end{tikzpicture}
    \caption{Circuit diagram for the teleportation based protocol that completes the $\mathbf{B}_{84}$ task. Blue lines indicate classical inputs and outputs. The protocol uses one EPR pair, $\ket{\Psi^+}=\frac{1}{\sqrt{2}}(\ket{00}+\ket{11})$. $b$ is a function of the classical measurement outcomes according to $(-1)^b= s_1^qs_2^{1-q}s_3$. While this is not the protocol employed by AdS/CFT, it illustrates how access to the central region $P$ can be replaced with use of entanglement.}
    \label{fig:BB84circuit}
\end{figure}

A linear bound is the best possible for this task, as there is a known way to complete $\mathbf{\hat{B}}_{84}$ by using $n$ EPR pairs, in which case $I/2=n$. This can be accomplished using the following protocol, first pointed out in \cite{lau2011insecurity}. Alice, upon receiving $H^q\ket{b}\in \mathcal{H}_{A_1}$, teleports the $A_1$ system using entanglement shared between near $c_1$ and near $c_2$. She obtains measurement outcomes $s_1$ and $s_2$. Near $c_2$, Alice applies $H^q$ to the teleported system, then measures in the computational basis, obtaining outcome $s_3$. We illustrate this protocol as a circuit diagram in figure \ref{fig:BB84circuit}. One can check that $b$ is a function of $s_1,s_2,s_3$ and $q$, specifically,
\begin{align}
    (-1)^b = s_1^q s_2^{1-q} s_3.
\end{align}
The measurement outcomes as well as $q$ may be copied and sent to both $r_1$ and $r_2$. Alice then computes $b$ according to the above relation near both $r_1$ and $r_2$ and outputs $\ket{b}$ as needed.

\section{Minimal surface and bulk central point calculations}\label{app:minsurfaceandbulkpoint}

We begin by recalling the minimal surface construction. In global coordinates on AdS$_{2+1}$, we have the metric
\begin{align} \label{eq:rcoords}
    ds^2 = -\left( 1+r^2 \right)dt^2 + \left(1+r^2\right)^{-1} dr^2 + r^2 d\varphi^2,
\end{align}
where we've measured lengths in units of the AdS length. Then the minimal surfaces are given by
\begin{align}
    r(\varphi) = \left(\frac{\cos^2\varphi}{\cos^2 \varphi_A}-1 \right)^{-1/2},
\end{align}
where $\varphi_A$ is the opening angle of the boundary interval the surface is anchored to. These geodesics have length
\begin{align}
    L = \log \left( \sin \left( \varphi_A \right) \right) + O(\log\epsilon ),
\end{align}
where $\epsilon$ is a cutoff in the integration range, which is taken over the interval $\phi \in (-\phi_A+\epsilon,\phi_A-\epsilon)$. 

We will consider the arrangement of regions shown in figure \ref{fig:min_surfaces}, where we have two regions of angular size $x$ separated by an angle $2\alpha$. The regions are located on the $t=0$ time slice. We may assume without loss of generality that $x/2+\alpha \leq \pi/2$, since otherwise we may measure the angle $\alpha$ from the other side of the disk. The condition for $I(R_1:R_2)>0$ from requiring the entanglement wedge to become connected is that
\begin{align}
    2 \log\sin(x/2) > \log \sin(\alpha+x) + \log \sin(\alpha),
\end{align}
where the divergent parts have cancelled. Rearranging we have
\begin{align}\label{eq:inequality}
    \sin^2(x/2) > \sin(\alpha+x)\sin(\alpha).
\end{align}
This is the condition we will compare to the one from quantum tasks. 

To find the condition from the quantum tasks perspective, it is more convenient to use the coordinates 
\begin{align}
    ds^2 = -\cosh^2(\rho) dt^2 + d\rho^2 + \sinh^2(\rho) d\phi^2
\end{align}
which are related to \ref{eq:rcoords} by the coordinate change
\begin{align}
    r = \sinh(\rho).
\end{align}
The boundary coordinates are unchanged. The first step is to work out the location of the points $c_1,c_2,r_1,r_2$ such that $D(R_i)=J^+(c_i)\cap J^-(r_1)\cap J^-(r_2)$. This is a straightforward exercise in (flat) Lorentzian geometry. Specifying the regions to be on the $t=0$ slice, the result is that
\begin{align}
    c_1 &= (-x/2,-\alpha-x/2) \nonumber \\
    c_2 &= (-x/2,\alpha+x/2) \nonumber \\
    r_1 &= (x+\alpha,0) \nonumber \\
    r_2 &= (\pi-\alpha,\pi).
\end{align}
Since in general it may be time-like or null curves that connect these four points to a central vertex, it is convenient to use only null rays but allow them to reach the boundary early, since we can always add a delay and have the ray reach the $r_i$'s exactly. 

\begin{figure}
    \centering
    \begin{subfigure}{0.45\textwidth}
    \begin{tikzpicture}
    
    \draw[thick] (0,0) circle (3);
    
    \draw[red,thick,postaction={on each segment={mid arrow}}] (-3,0) to [out=0,in=180] (3,0);
    \draw[red,thick,postaction={on each segment={mid arrow}}] (-3,0) to [out=30,in=150] (3,0);
    \draw[red,thick,postaction={on each segment={mid arrow}}] (-3,0) to [out=60,in=120] (3,0);
    \draw[red,thick,postaction={on each segment={mid arrow}}] (-3,0) to [out=-60,in=-120] (3,0);
    \draw[red,thick,postaction={on each segment={mid arrow}}] (-3,0) to [out=-30,in=-150] (3,0);
    
    \node[above] at (0,1.75) {$\ell>0$};
    \node[below] at (0,-1.75) {$\ell<0$};
    
    \end{tikzpicture}
    \caption{}
    \end{subfigure}
    \hfill
    \begin{subfigure}{0.45\textwidth}
    \begin{tikzpicture}
    
    \draw[thick] (0,0) circle (3);
    
    \draw[red,thick,postaction={on each segment={mid arrow}}] (0,-1) -- (0,-3);
    \draw[red,thick,postaction={on each segment={mid arrow}}] (0,-1) -- (0,3);
    
    \draw[red,thick,postaction={on each segment={mid arrow}}] (-2.12,-2.12) to [out=18.5,in=-144] (0,-1.05);
    
    \draw[red,thick,postaction={on each segment={mid arrow}}] (2.12,-2.12) to [out=161.5,in=-36] (0,-1.05);
    
    \draw[fill=black] (0,-1.05) circle (0.05);
    \node[above right] at (0,-1.05) {$p$};
    
    \draw[fill=black] (-2.12,-2.12) circle (0.05);
    \node[below left] at (-2.12,-2.12) {$c_1$};
    
    \draw[fill=black] (2.12,-2.12) circle (0.05);
    \node[below right] at (2.12,-2.12) {$c_2$};
    
    \draw[fill=black] (0,-3) circle (0.05);
    \node[above] at (0,3) {$r_2$};
    
    \draw[fill=black] (0,3) circle (0.05);
    \node[below] at (0,-3) {$r_1$};
    
    \end{tikzpicture}
    \caption{}
    \end{subfigure}
    \caption{a) Projection of null lines in pure global AdS onto a constant time slice. Null lines always begin and end at antipodal points, and their angular momentum $\ell$ controls their path. b) Projection of the minimal time paths connecting a central point $p$ with the input and output points. From symmetry, only $\ell$, the angular momentum of one of the incoming null lines, is left to optimize over.}
    \label{fig:topview}
\end{figure}
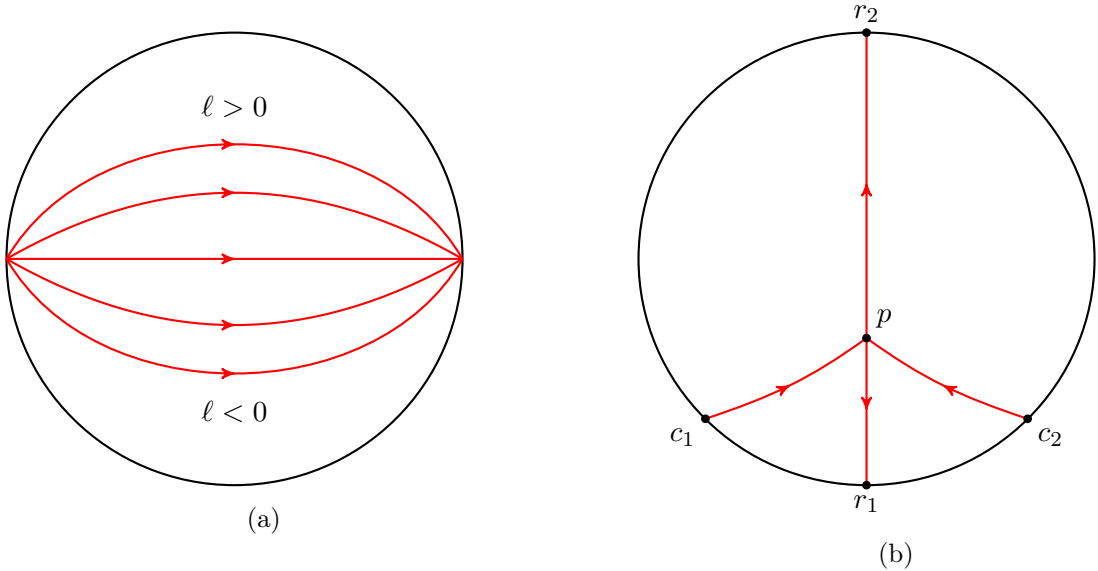

The shortest time paths are the null geodesics, which in parametric form are
\begin{align}
    \phi(\lambda) &= \phi_0 + \frac{\pi}{2} + \arctan\left( \frac{\lambda(1-\ell^2)}{\ell}\right) \nonumber \\
    \rho(\lambda) &= \text{arccosh}\left(\sqrt{\lambda^2(1-\ell^2) + \frac{1}{1-\ell^2}} \right) \nonumber \\
    t(\lambda) &= t_0 + \frac{\pi}{2} + \arctan \left( \lambda (1-\ell^2) \right).
\end{align}
$\phi_0,t_0$ are the starting angular location and time of the geodesic, while $\ell$ is the angular momentum. A top view of the null geodesics in vacuum AdS$_{2+1}$ is shown in figure \ref{fig:topview}. The angular momentum parameter $\ell$ lies in the range $(-1,1)$.  

Because we have limited ourselves to the case of intervals of equal size lying on a constant time slice, we can use the symmetry of our set-up to simplify our optimization problem. In particular we anticipate that the central point $p$, if it exists, will lie on the straight line that connects $\phi_{J1}$ and $\phi_{J2}$ corresponding to $\phi=0$. The two outgoing lines then will be $\ell=0$ rays from $p$ to $r_1$ and $p$ to $r_2$. Meanwhile, the two incoming lines must have equal and opposite angular momentum to enforce that they meet on the line connecting $r_1$ and $r_2$. 

Denote by $\Delta T[c_i \rightarrow p](\ell)$ the time it takes a null geodesic to reach $p$ from $c_i$. Notice that because of the symmetry in the set-up $\Delta T[c_1 \rightarrow p](\ell)=\Delta T[c_2\rightarrow p](\ell) \equiv \Delta T[c\rightarrow p](\ell)$. Call $\Delta T[p\rightarrow r_i](\ell)$ the time it takes to travel from $p$ to the spatial location of $r_i$. Then the existence of a suitable central point $p$ requires that
\begin{align}\label{eq:timeintervals}
    \Delta T[c\rightarrow p](\ell) + \Delta T[p\rightarrow r_1] &\leq \frac{3x}{2} + \alpha \\
    \Delta T[c\rightarrow p](\ell) + \Delta T[p\rightarrow r_2] &\leq \pi - \alpha + \frac{x}{2}.
\end{align}
The time intervals on the left hand side all depend on the geometry of the regions through $x,\alpha$. The time to reach $p$ from $c$ depends additionally on the angular momentum parameter $\ell$. We consider $x,\alpha$ as fixed and ask if there exists an $\ell$ such that the above inequalities are satisfied. 

Using the solution for the null geodesics we can find the time intervals appearing in \ref{eq:timeintervals} as a function of $\ell$, 
\begin{align}
    \Delta T[c\rightarrow p](\ell) &= \frac{\pi}{2} - \arctan\left(\ell \cot\left(\frac{x}{2}+\alpha\right)\right) \nonumber \\
    \Delta T[p\rightarrow r_1] &= \frac{\pi}{2} - \text{arcsec} \left( \cosh(\rho_*) \right) \nonumber \\
    \Delta T[p\rightarrow r_2] &= \frac{\pi}{2} + \text{arcsec} \left( \cosh(\rho_*) \right),
\end{align}
where $\rho_*$ is the radial coordinate of $p$, and is given by
\begin{align}
    \cosh \rho_* = \sqrt{\frac{1+\ell^2 \cot^2(x/2+\alpha)}{1-\ell^2}} .
\end{align}
It is convenient to rearrange the inequalities \ref{eq:timeintervals} to the form 
\begin{align}
    F[x,\alpha](\ell) \equiv 3x/2 + \alpha - \left(\Delta T[c\rightarrow p](\ell) + \Delta T[p\rightarrow r_1] \right) > 0 \nonumber \\
    G[x,\alpha](\ell) \equiv \pi - \alpha + x/2 - \left(\Delta T[c\rightarrow p](\ell) + \Delta T[p\rightarrow r_2] \right) >0.
\end{align}
We can check that there is a solution to $F[x,\alpha](\ell) = G[x,\alpha](\ell)$ for $-1\leq \ell \leq 1$ whenever $x/2+\alpha \leq \pi/2$, which we have by assumption. Then notice that $\Delta T[p\rightarrow r_1](\ell)$ is decreasing as a function of $\ell$ when $\ell>0$ and increasing when $\ell<0$. Meanwhile, $\Delta T[p\rightarrow r_2](\ell)$ is decreasing when $\ell < 0$ and increasing when $\ell>0$. This means that if there is a region in $\ell$ over which $F[x,\alpha](\ell)>0$ and $G[x,\alpha](\ell)>0$, then it will include the point $\ell_*$ such that $F[x,\alpha](\ell_*) = G[x,\alpha](\ell_*)$.

Setting $F[x,\alpha](\ell_*)=G[x,\alpha](\ell_*)$ and solving for $\ell_*$ we get
\begin{align}
    \ell_* = \pm \frac{1}{\sqrt{1+\sec^2(\alpha+x/2)}}.
\end{align}
We can then substitute this into $F$ (or $G$, since they are equal at $\ell_*$) to get the values of $F=G$ at the intersection points. Taking the larger of the two intersection points we get that
\begin{align}
    \cot^2(\alpha+x/2) \geq \cot^2(x) \left(1+\sec^2(\alpha+x/2) \right).
\end{align}
Application of trigonometric identities shows this is equivalent to the inequality \ref{eq:inequality}, which we had found from the minimal surface argument. 

\bibliographystyle{unsrt}
\bibliography{biblio.bib}

\end{document}